\documentclass[aps,prb,twocolumn,superscriptaddress,showpacs,10pt,amssymb,amsfonts,floatfix]{revtex4-1}
\usepackage{amsmath,latexsym,epsfig,bm}

\usepackage[usenames,dvipsnames]{color}
\usepackage[colorlinks,bookmarks=false,citecolor=blue,linkcolor=red,urlcolor=blue]{hyperref}

\def\beq{\begin{equation}}
\def\eeq{\end{equation}}
\def\be{\begin{equation}}
\def\ee{\end{equation}}

\def\ket#1{\vert #1 \rangle}

\def\me#1#2#3{\langle #1 \vert #2 \vert #3 \rangle}
\newcommand{\bfm}{\mathbf}
\def\d{\partial}

\def\tr{\hbox{tr}\,}

\def\rf#1{(\ref{#1})}
\def\rfs#1{Eq.~\rf{#1}}

\def\dbyd#1#2{\frac{d#1}{d#2}}
\def\pp#1{\frac{\partial}{\partial#1}}
\def\pbyp#1#2{\frac{\partial#1}{\partial#2}}

\def\down{\downarrow}
\def\up{\uparrow}

\begin{document}

\title{Topological invariants and interacting one-dimensional fermionic systems}

\author{Salvatore R. Manmana}
\affiliation{Department of Physics, CB390, University of Colorado,
Boulder CO 80309, USA}
\affiliation{JILA (University of Colorado and NIST), Boulder CO 80309, USA}
\author{Andrew M.~Essin}
\affiliation{Department of Physics, CB390, University of Colorado,
Boulder CO 80309, USA}
\author{Reinhard M. Noack}
\affiliation{Fachbereich Physik, Philipps-Universit\"at Marburg, 35032 Marburg, Germany}
\author{Victor~Gurarie}
\affiliation{Department of Physics, CB390, University of Colorado,
Boulder CO 80309, USA}
\date{\today}

\begin{abstract}
We study one-dimensional, interacting, gapped fermionic systems described by variants of the Peierls-Hubbard model, and characterize their phases via a topological invariant constructed out of their Green's functions. 
We demonstrate that the existence of topologically protected, zero-energy states at the boundaries of these systems can be tied to the value of the topological invariant, just like when working with the conventional, noninteracting topological insulators.
We use a combination of analytical methods and the numerical density matrix renormalization group method to calculate the values of the topological invariant throughout the phase diagrams of these systems, thus deducing when topologically protected boundary states are present. 
We are also able to study topological states in spin systems because, deep in the Mott insulating regime, these fermionic systems reduce to spin chains. 
In this way, we associate the zero-energy states at the end of an antiferromagnetic spin-one Heisenberg chain with a topological invariant equal to $2$. 
\end{abstract}

\maketitle

\section{Introduction}
\label{sec:introduction}
Topological insulators---\emph{free} fermionic systems with topological band structure\cite{Ryu2010, HasanKane2010, Moore2010, QiZhang2010, HasanMoore2011, QiZhang2010-2}---are now very well understood. 
A band structure can be called topological if it has a nonvanishing topological invariant, the Chern number first proposed for the integer quantum Hall effect being the simplest example.\cite{Thouless1982} 
These invariants imply zero-energy boundary states; it is these boundary states that distinguish the topological insulators from their non-topological counterparts and which are crucial for their physical properties. 
However, a number of topological states of matter have been discovered whose existence requires \emph{interactions}.   
Prominent among these are the ``topologically ordered'' states in two dimensions as defined by X.-G. Wen, inspired by the fractional quantum Hall effect.\cite{Wen1995} 
Another example is the Haldane state of the spin-one antiferromagnetic Heisenberg chain.\cite{Haldane1983} 
Like noninteracting topological insulators, these states are bulk-incompressible with zero-energy excitations at the boundary. 
However, they are substantially different from the topological insulators in the need for interactions, and their excitations are often fractionalized relative to the underlying, microscopic degrees of freedom as, e.g., Laughlin's fractionally charged excitations in the fractional quantum Hall effect and the spin-1/2 boundary states of the Haldane state of the spin-1 Heisenberg chain. 
It is therefore natural to ask whether there is a connection between topological band structures and interacting topological states.  

Here, we explore one connection proposed in the literature in recent years,\cite{VolovikBook1,Wang2010,zhong1,zhong2,Gurarie2011} where one computes topological invariants of the single-particle Green's function rather than the single-particle Hamiltonian.
These invariants coincide in the absence of interactions but, unlike single-particle Hamiltonians, single-particle Green's functions continue to exist even when interactions are present.  Their topological invariants thus
generalize the free invariants to generic, interacting systems.

In this paper, we use this approach to compute explicitly a topological invariant for one-dimensional, interacting, fermionic systems, and connect its value to the presence or absence of topologically protected, zero-energy boundary states.
In particular, we study Hubbard models with dimerization, as well as spin chains, which can be understood as Mott-insulating phases of the Hubbard models in the strongly interacting regime. 
Where possible, we calculate invariants analytically but, in the general case, we rely on the numerical density matrix renormalization group method\cite{white1992,white1993,Schollwock:2005p2117} and its time-dependent extension (adaptive t-DMRG\cite{Daley:2004p2943,White:2004p2941}) to compute the invariant as a function of the system parameters. 
Being an integer, the invariant is not very sensitive to the errors inherent in our numerical approach and so can be determined precisely with moderate computational effort.
The ultimate goal of this paper is not so much to calculate the phase diagram of the models we study as to illustrate the utility of the Green's-function method of topological invariants when applied to interacting, one-dimensional, fermionic systems. 

While we find this approach to be useful, some caution is needed in the interpretation of these interacting topological invariants.  First, in the absence of interactions the topological invariants frequently measure the linear response to external (frequently electromagnetic) perturbations.\cite{Thouless1982,Ludwig2010}  For example, the Chern number characterizing the integer quantum Hall states is proportional to the Hall conductance $\sigma_{xy}$ of these systems.  However, this connection is not guaranteed for the Green's function invariant of an interacting system.
Second, interactions introduce a novel possibility for the physics at the boundary not present in free systems.  In particular, in the free system, a nontrivial value for the bulk invariant implies zero-energy (single particle) excitations at the boundary, which formally appear as poles of the single-particle Green's function.  In an interacting system, a nontrivial bulk invariant is also consistent with zero-energy \emph{zeros} of the Green's function,\cite{Gurarie2011} in addition to poles,\cite{VolovikBook1,Essin2011} which indicates a complete loss of coherence of the single-particle degrees of freedom.  We will see this explicitly in examples.

In the bulk, this behavior --- a singularity of the Green's function without a corresponding zero-energy single-particle state --- means that the topological invariant can change its value (discontinuously) as parameters are varied without passing through a phase transition.  This means that, unlike for free systems, these topological invariants do not necessarily correspond in a simple way to phases of matter.
Fortunately, recent works have addressed this issue for one-dimensional fermion systems.\cite{Fidkowski2010,Fidkowski2011,Turner2011}  These authors find that there are \emph{only} four topologically distinct phases, characterized by the value of the topological invariant modulo 4.  [Note that the authors of these papers also consider fermionic systems for which the number of particles is not conserved, i.e., without U(1) symmetry, which leads to 8 distinct phases, with ${\mathbb Z}_8$ structure.  In this paper we restrict ourselves to systems with U(1) symmetry and ${\mathbb Z}_4$ structure.]  Given this knowledge, our computations of topological invariants provide unambiguous determination of the topological phase for each model we consider.

We apply these ideas first to the one-dimensional Peierls-Hubbard model of spin-1/2 fermions with dimerized hopping and on-site Hubbard repulsion (see, e.g., Ref.~\onlinecite{PhysRevB.66.045114}).  
Its invariant can take on the values 0 or 2.  
We also briefly treat a system of two coupled Hubbard chains, which realizes the Fidkowski-Kitaev model\cite{Fidkowski2010} and whose invariant can take on values which are multiples of four; i.e., this model has only one phase. 
We calculate the interacting invariants for these models using both analytical arguments and the t-DMRG method. 
We point out that the existence of the boundary states in these models can indeed be captured by these invariants, in accordance with the bulk-boundary correspondence.\cite{Essin2011} 
We further point out that when the invariant is a multiple of four, the disappearance of the boundary states is consistent with the bulk-boundary correspondence, thanks to the replacement of zero-energy boundary states by zeros in the Green's function.  
Finally, we analyze numerically and with some analytical arguments a variant of the fermionic Peierls-Hubbard chain whose parameters are adjusted so that the system is in a spin-one Haldane phase.
We find that the interacting topological invariant is equal to 2, showing that the spin-one Heisenberg chain possesses the same boundary states as the Peierls-Hubbard model.

The rest of the paper is organized as follows. In Sec.~\ref{sec:topinv}, we introduce the topological invariant for interacting one-dimensional systems. In Sec.~\ref{sec:hubbard}, we introduce the one-dimensional Peierls-Hubbard model and discuss its phase diagram, topological invariant, and boundary states analytically. In Sec.~\ref{sec:dmrg}, we complete the study of the phase diagram of the Peierls-Hubbard model and the topological invariant using the DMRG method. In addition, we analyze the boundary states in a variant of the Peierls-Hubbard model with spin interactions that is equivalent to the spin-one Haldane chain. 
In Sec.~\ref{sec:fk}, we show that the absence of phase transitions in the Fidkowski-Kitaev model is compatible with a changing invariant precisely because the Green's function acquires zeros at a certain point of the phase diagram, leading to a change of the invariant by a multiple of 4. Finally, in Sec.~\ref{sec:conclusions}, we present our conclusions and outlook.

\section{One-dimensional topological insulators and their topological invariant}
\label{sec:topinv}
Consider a one-dimensional fermionic system. Two types of topological invariants are known to exist for these systems (in the absence of interactions), one with ${\mathbb Z}$ structure and
one with ${\mathbb Z}_2$ structure. In this paper we concentrate on those systems whose invariant is an integer ${\mathbb Z}$. To allow for the existence of the topological invariant, the imaginary-time single-particle Green's function $G(k,\omega)$ of such systems must possess the symmetry \cite{Gurarie2011,Foster2006}
\be \label{eq:chi} 
\Sigma G(k,\omega) \Sigma = - G(k,-\omega),
\ee
where $\Sigma$ is some unitary matrix whose square is 1. 
Here $\omega$ is the imaginary frequency and $k$ is the wave vector. 
This symmetry is usually referred to as a chiral symmetry.\cite{Ryu2010,Gurarie2011} 
In the absence of interactions, it occurs when a particle moves on a bipartite lattice; in one dimension, this symmetry is present for a tight-binding model with nearest-neighbor hopping only.  
In the presence of interactions, it appears as a combination of particle-hole and time-reversal transformations and is also very ubiquitous. In particular, adding Hubbard-type interactions to the tight-binding model with sublattice symmetry preserves \rfs{eq:chi}. 

A Green's function with the property \rfs{eq:chi} is characterized by a topological invariant. Defining 
\be \label{eq:defy} 
g(k) = \left. G(k,\omega) \right|_{\omega=0},
\ee 
we can write the invariant as\cite{Volovik2010,zhong1,zhong2}
\be \label{eq:top} 
N_1 = \tr \, \int \frac{dk}{4\pi i} \, \Sigma g^{-1} \d_k g.
\ee
Here the trace is taken over the matrix indices of $g$ (which label the bands and spin indices of the model we study). Written in this form, this invariant exists 
whether or not interactions are present. One comment is in order: this definition of the topological invariant differs from the one adopted in Ref.~\onlinecite{Fidkowski2010} by a factor of 2, since, as was pointed out in Sec.~\ref{sec:introduction}, we work with complex (or Dirac) fermions with a  conserved number of particles, while Ref.~\onlinecite{Fidkowski2010}  works with real (or Majorana) fermions. The subscript $1$ in $N_1$ refers to the one-dimensional space in which it is defined. 

It is straightforward to see that $N_1$ is topological, that is, it does not change if one changes $g(k)$ slightly. Indeed, if $g$ depends on some parameter $\alpha$ (a coupling constant in the Hamiltonian, for example), the derivative $dN_1/d\alpha$ can be found to be 
\be \label{eq:stability} 
\dbyd{N_1}{\alpha} = \tr \, \int \frac{dk}{4\pi i} \, \Sigma \, \d_k \left( g^{-1} \d_\alpha g \right) =0, 
\ee 
i.e., as an integral over a total derivative. 
To show that \rfs{eq:stability} holds, one must take advantage of \rfs{eq:chi}, which implies, together with \rfs{eq:defy}, that $g$ anticommutes with $\Sigma$,
\be \label{eq:chi1} 
\Sigma g = - \Sigma g.
\ee
It is well-known that a basis always exists for which the matrix $\Sigma$ takes the form
\be \label{eq:sigmaexpl} \Sigma = \left( \begin{matrix} 1 & 0 \cr 0 & -1 \end{matrix} \right), 
\ee
where $1$ stands for an identity matrix. 
This allows us to rewrite the topological invariant, \rfs{eq:top}, in a slightly different form by noting that 
thanks to the condition (\ref{eq:chi1}) as well as to \rfs{eq:sigmaexpl}, the matrix $g$ must have the off-diagonal structure
\be \label{eq:offdiag} g = \left( \begin{matrix} 0 & v(k) \cr v^\dagger(k) & 0 \end{matrix} \right),
\ee
where $v(k)$ is some generic matrix. Substituting \rfs{eq:offdiag} into \rfs{eq:top}, we find
\be \label{eq:winding} N_1 =\tr \, \int \frac{dk}{2\pi i} \, \partial_k \log v(k) = \sum_n \int \frac{dk}{2\pi i} \partial_k \log z_n(k),
\ee
where $z_n(k)$ are the eigenvalues of $v(k)$. We see that $N_1$ simply counts the number of eigenvalues of $v$ that  wind around the origin of the complex plane as a function of $k$. 

In practical applications below, $v(k)$ will often be just a number or a diagonal matrix, and $z_n(k)$ will be very straightforward to identify.     

In the absence of interactions, it is possible to relate $G$ and $g$ directly to the Hamiltonian. Indeed, a generic noninteracting Hamiltonian looks like
\be \hat H = \sum_{\alpha \beta} {\cal H}_{\alpha \beta} \hat c^\dagger_{\alpha} \hat c_\beta,
\ee where $\hat c$ and $\hat c^\dagger$ are annihilation and creation operators, and the indices $\alpha$, $\beta$ refer to lattice sites, spin, and flavor of fermions, if any. Its Green's function is given by
\be \label{eq:noint} G = \left[ i \omega - {\cal H} \right]^{-1}, \ g=-{\cal H}^{-1}.
\ee 
With this identification, the invariant, Eq.~(\ref{eq:top}), becomes the well-known one-dimensional version of the topological invariant for noninteracting fermionic systems with chiral symmetry.\cite{Ryu2010} 
It is used, for example, to identify topological phases  and boundary states of fermionic chains such as those studied in Ref.~\onlinecite{Su1979} (whose boundary states are often referred to as Su-Schrieffer-Heeger solitons). 

Once interactions are turned on, however, simple expressions such as \rfs{eq:noint} are no longer available.  
The utility of the topological invariant \rfs{eq:top} lies in the following. First, in the absence of interactions, the only way for $N_1$ to change is if $g$ becomes singular at some momentum $k$.
This can happen only if the system has zero-energy excitations as follows from \rfs{eq:noint}, thus implying a quantum phase transition. In the presence of interactions, $N_1$ can also change if $g$ acquires zero eigenvalues at some $k$. (This is impossible in the absence of interactions as follows from \rfs{eq:noint} assuming that the Hamiltonian is bounded; see also Ref.~\onlinecite{Silaev2012}, which relaxes this assumption.) This is the origin of the possibility that $N_1$ may change value even in the absence of a quantum phase transition. 

Second, suppose we have two adjacent domains where the invariant $N_1$ takes on two different values, $N_1^R$ in the right domain and $N_1^L$ in the left one. Then one can show that\cite{Essin2011}
\be \label{eq:bb} N_1^R - N_1^L = \tr \Sigma , 
\ee
where the trace is evaluated in the Hilbert space of the chain with an open 
boundary.
This constitutes the bulk-boundary correspondence for this type of topological insulator, and $\tr  \Sigma$ can be viewed as a boundary topological invariant, as we will see below. The derivation of \rfs{eq:bb} (see, in particular, Appendix B of Ref.~\onlinecite{Essin2011}) is based on the algebraic manipulations of the function $g$, independent of its physical meaning (in particular, independent of whether interactions are present). 

To see the implication of this relation for the boundary states, we calculate this trace in the basis of the eigenstates of $g_{\alpha \beta}$, the zero-frequency Green's function of the open chain (which is thus not translationally invariant and so $g$ cannot be reduced to just a function of the momentum $k$). Every eigenstate $\psi_n$ of $g$ with nonzero eigenvalue $\lambda_n$,
\be g \, \psi_n = \lambda_n \psi_n, \ee
has a conjugate eigenstate with an opposite eigenvalue,
\be g \, \Sigma \psi_n = - \lambda_n \Sigma \psi_n ,
\ee
as a consequence of \rfs{eq:chi1}. It follows that 
\be \psi_n^* \Sigma  \psi_n=0\ee 
because $\psi_n$ and $\Sigma \psi_n$ are both eigenstates of $g$ with opposite eigenvalues. Thus eigenstates with nonzero eigenvalues $\lambda_n$ do not contribute to $\tr \Sigma$. Only the eigenstates of $g$ with $\lambda_n$ either infinite (poles of the Green's function at $\omega=0$) or zero (zeros of the Green's function at $\omega=0$) contribute to the trace. So we see that $\tr \Sigma$ must be counting zero-energy states present on the boundary between two topological insulators, or possibly zeros of Green's functions if there are interactions, justifying its earlier characterization as a boundary topological invariant. 

Moreover, all such eigenstates with $\lambda_n$ either infinite or zero are also eigenstates of $\Sigma$ with eigenvalues that are either $+1$ or $-1$. 
To see this, we observe that if $\psi_n$ is an eigenstate with a zero eigenvalue, then so is $\Sigma \psi_n$. 
By forming linear combinations $\psi_n \pm \Sigma \psi_n$, we construct the eigenstates that are also eigenstates of $\Sigma$ with the promised eigenvalues. 
Similar arguments can be given for eigenstates with infinite $\lambda_n$. 
Thus, $\tr \Sigma$ counts these types of eigenstates with appropriate signs. 
We see that \rfs{eq:bb} simply tells us that a boundary between two topological insulators with different values of the topological invariant must have either some zero energy states, or some zeros of the Green's function, or perhaps both. 

We will see that in practical applications of \rfs{eq:bb} a third possibility can arise. 
Namely, the existence of zero-energy excitations signifies that the ground state is degenerate and the system must pick one of the degenerate states spontaneously. 
The ground state chosen in this way may break the chiral symmetry and \rfs{eq:chi} may then break down. 
In this case, it may happen that the system has no single-particle zero-energy excitations and no zeros of the Green's function, since \rfs{eq:bb} is simply no longer valid. 
This happens in the dimerized Hubbard model studied in the next section under certain conditions. 
However, the system still has zero-energy boundary states, which is reflected in the existence of more than one ground state. 
It is just that the Green's function, which is sensitive to the single-particle excitations only, does not see those zero-energy excitations corresponding to the multiple ground states, which are actually of  particle-hole type in this case. The main conclusion remains the same: under these conditions a nonzero bulk topological invariant implies zero-energy states at the edge. 

Finally, as we have pointed out in Sec.~\ref{sec:introduction}, at the boundary of a topological system with the invariant $N_1$ equal to 0 modulo 4, there should be no zero-energy states. 
That means that, instead, there will only be zeros, in accordance with \rfs{eq:bb}. The same is true for the boundary between two insulators with the difference of $N_1$ across the boundary equal to 0 modulo 4. The converse is also true: if the invariant changes by a number other than 0 (mod 4) across a boundary, there will always be some zero-energy states. This follows from the arguments of Refs.~\onlinecite{Fidkowski2010,Fidkowski2011,Turner2011} as well as for the reasons presented above.  One interesting possibility is that the boundary states are not of single-particle nature and that the Green's function does not actually have any poles at $\omega=0$, but that there are zero-energy states that are collective excitations of many particles. 
In this case, the Green's functions will have zeros at the boundary to satisfy \rfs{eq:bb}, but the actual system will still have zero-energy excitations. (However, we have not observed such a scenario in any of the examples considered in this paper). 

Regardless of the mechanism, the main conclusion is that if $N_1$ is equal to $1$, $2$, or $3$ modulo 4 (or if it changes by this amount across the boundary between two insulators), there must be zero-energy states of some kind at the edge of the system.

\section{One-dimensional Peierls-Hubbard model}
\label{sec:hubbard}
Now let us consider a model of spin-1/2 fermions moving in one dimension on a line with dimerized hoppings and on-site interactions, at half filling (one fermion per site). The Hamiltonian is given by 
\begin{eqnarray} \label{eq:ham}
\hat H &=& 
\sum_{j, \sigma=\uparrow, \downarrow} \left[ t-(-1)^j \delta t \right] 
\left( \hat c^\dagger_{j+1,\sigma} \hat c_{j,\sigma} +  \hat c^\dagger_{j,\sigma} \hat c_{j+1,\sigma} \right) \cr
&& +\,  U\,  \sum_j  \left( \hat n_{j,\uparrow} - \frac 1 2\right) \left( \hat n_{j,\downarrow} - \frac 12 \right), \cr  
{\rm where} && \\
\hat n_{j,\sigma} &=& \hat c^\dagger_{j, \sigma} \hat c_{j,\sigma} 
\end{eqnarray}
and $\hat{c}^\dagger_{j,\sigma}$ creates a particle of spin $\sigma$ on site $j=1\dots L$.  The ground state of this Hamiltonian is at half filling, that is, at one particle per site.  This can be verified by observing that 
this Hamiltonian is invariant under a properly defined particle-hole transformation (see appropriate detailed discussions in Ref.~\onlinecite{Gurarie2011})
\be 
\hat c^\dagger_{j,\sigma} \rightarrow (-1)^j \hat c_{j,\sigma}, \, \hat c_{j,\sigma} \rightarrow \hat c^\dagger_{j,\sigma} (-1)^j.
\ee
Chiral transformations are particle-hole transformations combined with the time-reversal operation. The Hamiltonian \rfs{eq:ham} is time-reversal invariant, so it is also chirally invariant. As a consequence of chiral symmetry, the Green's function satisfies \rfs{eq:chi}. Chiral symmetry may still be broken spontaneously by the ground state, in which case the Green's function would then violate \rfs{eq:chi} despite the Hamiltonian's invariance. This observation will be important later on. 

For what follows we need an explicit form of the generator of the chiral transformation. One can check that it is given by
\be \label{eq:sigmaop} 
\hat \Sigma = \prod_j 
\left( \hat c^\dagger_{j,\uparrow} +(-1)^j \hat c_{j,\uparrow} \right) 
\left( \hat c^\dagger_{j,\downarrow} +(-1)^j \hat c_{j,\downarrow} \right),
\ee
which is both hermitian and unitary. 
It is straightforward to see that
\be 
\hat \Sigma \hat c^\dagger_{j, \sigma} \hat \Sigma 
= (-1)^j \hat c_{j,\sigma}, \quad \hat \Sigma \hat c_{j, \sigma} \hat \Sigma 
= (-1)^j \hat c^\dagger_{j,\sigma},
\ee
and that 
\be \hat \Sigma \hat H \hat \Sigma = \hat H,
\ee
which expresses the fact that the Hamiltonian \rfs{eq:ham} is chirally symmetic.

Chiral invariance together with time-reversal (and spin-rotation) invariance places this Hamiltonian in the symmetry class BDI,\cite{Ryu2010} although one should remark that normally the concept of symmetry classes is applied to noninteracting systems only. But together with our expanded definition of topological invariants, we can extend this classification to interacting systems. 

In the absence of interactions, i.e., when $U=0$, this Hamiltonian describes a particle hopping on the lattice with staggered hopping.  Its topological invariant is straightforward to compute. Indeed, the Hamiltonian reduces to a matrix in this case,
\be \label{eq:nonint} \hat H = \sum_{j j',\sigma} {\cal H}_{j j'} \hat c^\dagger_{j, \sigma} \hat c_{j',\sigma}.
\ee
To write this matrix in momentum space, we observe that the unit cell for this chain consists of two sites.
By comparing \rfs{eq:nonint} to \rfs{eq:ham} at $U=0$ we deduce that
\be \label{eq:unitcell} 
{\cal H}(k) = 
\left( \begin{matrix}  
0 & t + \delta t + \left( t - \delta t \right) e^{ik} \cr  
t + \delta t + \left( t - \delta t \right) e^{-ik} & 0 
\end{matrix} \right).
\ee
Here the lattice spacing is set to 1, and $k\in (-\pi,\pi]$.  In the same basis, the matrix $\Sigma$ takes its standard  form
\be \label{eq:sigma} \Sigma = \left( \begin{matrix} 1 & 0 \cr 0 & -1 \end{matrix} \right).
\ee
By means of arguments presented earlier and given by Eqs.~\rf{eq:offdiag}, \rf{eq:winding} and \rf{eq:noint},
$N_1$ simply measures the winding of the upper right corner \cite{Ryu2010} of ${\cal H}$ around the origin of the complex plane as $k$ goes from $-\pi$ to $\pi$. 
This can be seen by substituting $g=-{\cal H}^{-1}$ into \rfs{eq:top} and observing that
\be 
N_1 = 2 \int \frac{dk}{2\pi i} \, \pp{k} \log z,
\ee 
where $z= t+ \delta t + \left(  t - \delta t \right) e^{ik}$. 
The factor of $2$ in front of the integral signifies the fact that there are two identical copies of $g$, one for each spin. 
Finally, one immediately sees that
\be 
\label{eq:invint} 
N_1 = \begin{cases} 2, & {\rm for}\ \delta t <0 \\ 0, & {\rm for}\ \delta t >0 \end{cases} .
\ee
We have thus reproduced well-known facts about dimerized, noninteracting chains. In particular we see that they have a topological phase transition if $\delta t$ is tuned to be zero. If $\delta t$ is made space-dependent so that
$\delta t>0$ to the right of a given point in space  and $\delta t<0$ to the left of that point, there are zero-energy states localized in the vicinity of that point (as mentioned earlier, called Su-Schrieffer-Heeger solitons 
\cite{Su1979} in this context).

Now let us examine Hamiltonian (\ref{eq:ham}) when $U>0$. 
First take $\delta t = 0$. In this case, this model is also very well understood. 
It is a Mott insulator, with a charge gap and gapless spin excitations described by a spin-1/2 antiferromagnetic chain.\cite{Lieb1968}  
The strength of antiferromagnetic couplings between nearby spins is $\sim t^2/U$ if $U \gg t$. However, $\delta t \neq 0$ dimerizes the spin-1/2 chain by making its bonds alternate in strength (while the charge sector remains a Mott insulator). 
From the theory of spin-1/2 chains it is known that a perturbation consisting of nearest-neighbor spin-spin interactions with alternating sign is relevant.\cite{Affleck1989}  
From this discussion, it is natural to expect that even for $U>0$ the system is gapped for all $\delta t$ except $\delta t =0$, where a quantum phase transition occurs. This parallels the case without interactions. 

\begin{figure}[t]
\includegraphics[width=\columnwidth]{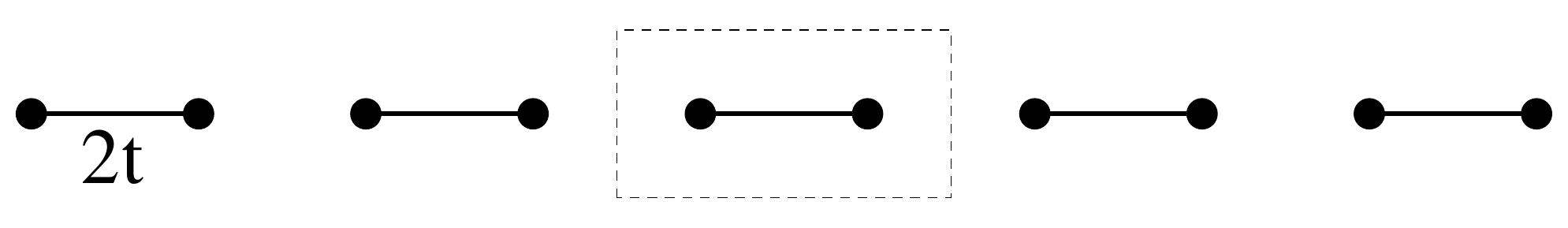}
\caption{\label{fig:dimer1}
A fully dimerized finite Hubbard chain with $|\delta t| = t$ consists of bonds of strength $2t$ (solid lines) and bonds of strength zero (not shown). The dashed box depicts the unit cell. In the case shown here, the topological invariant is $0$ and there are no boundary states.}
\end{figure}
\begin{figure}[b]
\includegraphics[width=\columnwidth]{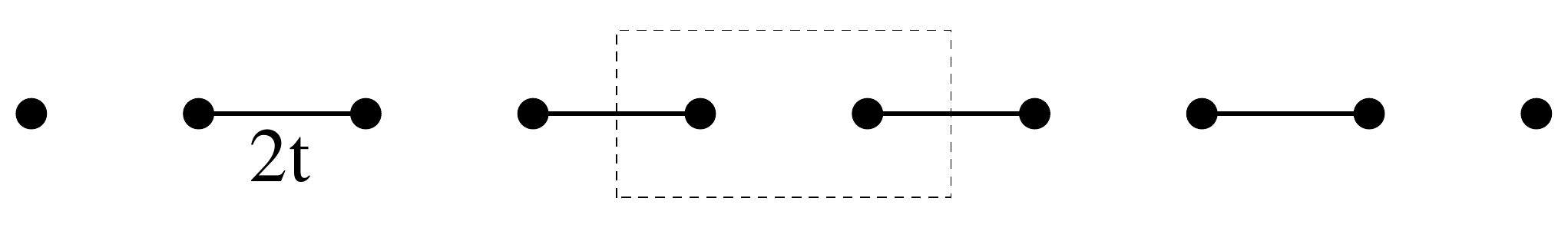}
\caption{\label{fig:dimer2}
A fully dimerized finite Hubbard chain as in Fig.~\ref{fig:dimer1}, but with the alternate dimerization pattern; again, the unit cell is depicted by the dashed box. In the case shown here, the topological invariant is $2$ and there are boundary states located at the isolated sites at both ends of the chain.}
\end{figure}

To understand the behavior of the topological invariant for $U>0$, let us consider a limiting case when $\delta t = \pm t$.
The chain then breaks up into disconnected clusters, each consisting of two sites only. The Green's function under these conditions can be found analytically by solving the two-site problem directly. However, it is not even necessary to solve the two-site problem to find the topological invariant. Indeed, the Green's function $G_{j j'}$ of this problem is zero unless both $j$ and $j'$ belong to the same two-site cluster. Its Fourier transform depends on whether a unit cell chosen previously when computing \rfs{eq:unitcell} coincides with the cluster, or if in a given connected cluster one site belongs to one unit cell and the other site to an adjacent unit cell. The two cases are distinguished by the sign of $\delta t$. In the former case, illustrated in Fig.~\ref{fig:dimer1}, the Green's function is momentum-independent. Then $g$ is also momentum-independent and $N_1$ is obviously zero. In the latter case, illustrated in Fig.~\ref{fig:dimer2}, the Green's function at zero frequency has the following structure in momentum space:
\be \label{eq:gtt} 
g(k) = 
\left( \begin{matrix} 0 & g_{12} \, e^{ik} \cr 
g_{12}^* \, e^{-ik} & 0 \end{matrix} \right).
\ee
Indeed, we know that the matrix $g$ must have zeros on its diagonal because $g$ anticommutes with $\Sigma$, and it is easy to see that $g_{12}$ must be a momentum-independent constant. 
Regardless of the actual value of $g_{12}$, as long as the Green's function has neither poles nor zeros at $\omega=0$ (that is, as long as $g_{12}$ is neither infinity nor zero), the topological invariant in this case is $N_1=2$, as can be established by substituting Eqs.~\rf{eq:gtt} and \rf{eq:sigma} into \rfs{eq:top}.

Therefore, we have established that \rfs{eq:invint} still applies even when $U>0$ in the extreme case $\delta t=\pm t$.
Earlier we saw that it also applies for all $\delta t$ if $U=0$. 
Since we expect this model to be gapped, with the line $\delta t = 0$ in the $U$ vs $\delta t$ phase diagram gapless, it is natural to conjecture that \rfs{eq:invint} applies for all $\delta t$ and all $U$. We will verify this in the next section. 
For now let us explore what knowledge of the value of the topological invariant implies for the boundary states of this insulator. 

According to the discussion in the previous section, we expect that a phase with  $N_1=2$ has either zero energy states or zeros of the Green's function (or both) at its boundary. We would like to understand which of these possibilities is realized in our case.  To do that, let us reexamine  the case when $\delta t= \pm t$. 

When the unit cell consists of two sites at the ends of a nonzero bond, there are clearly no boundary states, as can be seen in Fig.~\ref{fig:dimer1} (and in that case, as we just saw, $N_1=0$).  
In the other case, when $N_1=2$, there will be a single unpaired site at each end of the chain (shown in Fig.~\ref{fig:dimer2}). The Green's function for that site is straightforward to compute. The Hamiltonian of that single site is
\be 
\hat{H}_{j_0} = U \, \left( \hat c^\dagger_{j_0,\uparrow} \hat c_{j_0, \uparrow}  - \frac 1 2 \right) \left( \hat c^\dagger_{j_0,\downarrow} \hat c_{j_0,\downarrow} - \frac 1 2 \right),
\ee
where $j_0=1$ or $j_0=L$  is the unpaired site at either end of the chain ($L$ is the total length of the chain). 
The Hamiltonian acts in the space of four states --- an empty site, a site occupied by a spin-up particle, a site with a spin-down particle,  and a site filled with two particles --- with energies $U/4$, $-U/4$, $-U/4$, and $U/4$, respectively. The ground state is then a site with either a spin-up or a spin-down fermion. These two states are degenerate. 
Therefore, it tells us that the ground state of the entire one-dimensional chain with two ends is four-fold degenerate. 
These are what we can call the boundary states in the many-body context. 

Let us see how this is reflected in the Green's function. 
Choosing a state with either a spin-up or a spin-down particle as the ground state, we can
calculate the Green's function directly from its spectral (Lehmann) decomposition. 
We find that the Green's function at this site is
\be \label{eq:greensite} G_{j_0, \sigma; j_0, \sigma'} = \left( \begin{matrix}  \frac{1}{i \omega - U/2} & 0 \cr 0 & \frac{1}{i\omega + U/2} \end{matrix} \right)_{\sigma,\sigma'}.
\ee
Here the $2\times2$ structure of the Green's function is in spin space, with the first row and the first column of the matrix corresponding to the spin-up state  if the ground state is chosen to be the state filled with the spin-down fermion, or vice versa. 

We see that the Green's function has neither poles nor zeros at $\omega=0$. The absence of poles is not surprising: the system is a Mott insulator and adding or removing a particle should cost a finite energy, $U/2$ in this case. Yet  we need to see that this is compatible with, and indeed follows, from the bulk-boundary correspondence, \rfs{eq:bb}. 
At first glance, it contradicts the correspondence since we have a state with the topological invariant, $N_1=2$, but without zeros or poles of the Green's function at $\omega=0$.

Looking more closely, however, we observe that having a state with one particle on a site with a given spin violates particle-hole symmetry. Indeed, the state
$\hat c^\dagger_{j_0, \uparrow} \left| 0 \right>$ goes into $\hat  c^\dagger_{j_0,\downarrow} \left|0 \right>$ under the particle-hole transformation defined in \rfs{eq:sigmaop}, and vice versa; the doublet is symmetric, not the states themselves. Mathematically, this is expressed in the fact that \rfs{eq:greensite} does not satisfy \rfs{eq:chi}. In other words, we observe that the presence of the many-body boundary states is reflected in the spontaneous breaking of the particle-hole (and chiral) symmetry. 

To summarize, we find that $N_1=2$ does not result in either poles or zeros of the Green's function. This is not compatible with \rfs{eq:bb}, as long as \rfs{eq:chi} holds. The way out is to recognize that the violation of \rfs{eq:bb} is only possible if there is a spontaneous breaking of the chiral symmetry and \rfs{eq:chi} is violated. 
In turn, this means that the ground state is multiply degenerate, which is a signature of the presence of zero-energy boundary states. 
This is how we can reconcile the value $N_1 = 2$ with the absence of
zero-energy, \emph{single-particle} boundary states. 

But what if $\delta t$ is close but not quite equal to $\pm t$ and the chain is short enough that the boundary states can entangle and restore the chiral symmetry? 
Concentrating on the two ends of the chain with $j_0=1$ and $j_0=L$ with $L$ the length of the chain, we find that the ground state must take the form
\be 
\left( 
\alpha \hat c^\dagger_{1,\uparrow} \hat c^\dagger_{L, \uparrow} 
- \alpha^*  \hat c^\dagger_{1,\downarrow} \hat c^\dagger_{L, \downarrow} 
+ \beta \hat c^\dagger_{1,\uparrow} \hat c^\dagger_{L, \downarrow} 
+ \beta^*  \hat c^\dagger_{1,\downarrow} \hat c\dagger_{L, \uparrow} 
\right) \left| 0 \right>, 
\ee
with arbitrary amplitudes $\alpha$ and $\beta$ such that $|\alpha|^2 +|\beta|^2=1$. 
Indeed, one can check that this state is an eigenstate of $\hat \Sigma$ using \rfs{eq:sigmaop}, which for these two sites reduces to
\beq 
\hat \Sigma = 
[ \hat c^\dagger_{1,\uparrow} -\hat c_{1,\uparrow} ]
[ \hat c^\dagger_{1,\downarrow} -\hat c_{1,\downarrow} ]
[ \hat c^\dagger_{L,\uparrow} +\hat c_{L,\uparrow} ]
[ \hat c^\dagger_{L,\downarrow} +\hat c_{L,\downarrow} ] .
\eeq
Here we ignore all the sites in the bulk, concentrating just on the two disconnected edge sites. 

Calculating the Green's function via its spectral decomposition yields 
\be G_{j_0,\sigma;j_0, \sigma'} = \delta_{\sigma \sigma'} \left(\frac{1}{i\omega-U/2} + \frac{1}{i \omega+U/2} \right).
\ee
This function is indeed chirally invariant, satisfying \rfs{eq:chi}. 
At the same time, it vanishes at $\omega=0$.  

We see that if the chiral symmetry is restored by considering a finite chain with a unique ground state, then the bulk-boundary relation, \rfs{eq:bb}, is compatible with $N_1=2$ by way of the Green's functions having zeros at the edges of the chain. 
Even though this implies that an $N_1=2$ chain does not have boundary states (something which would have been impossible in conventional topological chains without interactions), this lack of boundary states is an effect of the finite length, and the boundary states are restored when the length of the chain is taken to infinity. 

Finally, while we have only looked at $U>0$ so far, a well-known mapping takes the half-filled Hubbard model at positive $U$ to one at negative $U$. 
This is achieved by transforming 
\be 
\hat c^\dagger_{j,\uparrow} \rightarrow (-1)^j \hat c_{j,\uparrow}, \quad
\hat c_{j,\uparrow} \rightarrow (-1)^j \hat c^\dagger_{j,\uparrow},
\label{eq:mappingnegativeU}
\ee 
without changing $\hat c_\downarrow$ and $\hat c^\dagger_\downarrow$.  
Therefore, the whole discussion thus far applies equally well to the $U<0$ region of the phase diagram.

Let us now summarize what we have learned by the direct analysis of the case $\delta t = \pm t$. If the topological invariant is nonzero, one of three scenarios is realized. The first scenario is that the Green's function has poles at zero frequency, indicating zero-energy single-particle excitations. 
This occurs only at $U=0$ in our model. 
For $U>0$ the system is a Mott insulator, and adding a particle always costs finite energy. The boundary states, if present, are particle-hole excitations, not single-particle ones; the single-particle Green's function cannot detect them and remains gapped. Nevertheless, either the Green's function acquires zeros at the boundary or the chiral symmetry is spontaneously broken by the boundary. 
These possibilities constitute the second and the third possible scenario, respectively. 
Either result signifies the presence of multiparticle, zero-energy boundary states in the limit of an infinite chain. 

Finally, we note that the ground state of the Peierls-Hubbard model is closely connected to that of a spin-one Heisenberg chain. To elucidate this point, let us add a ferromagnetic term to the Peierls-Hubbard Hamiltonian on every other bond. The combined Hamiltonian reads

\begin{eqnarray} \label{eq:hamH}
\hat H &=& 
\sum_{j, \sigma=\uparrow, \downarrow} \left[ t-\delta t (-1)^j \right] 
\left( \hat c^\dagger_{j+1,\sigma} \hat c_{j,\sigma} +  \hat c^\dagger_{j,\sigma} \hat c_{j+1,\sigma} \right) + \cr
&& +\,  U\,  \sum_j  \left( \hat n_{j,\uparrow} - \frac 1 2\right) \left( \hat n_{j,\downarrow} - \frac 12 \right)+  \cr && J \sum_{j} {\bfm S}_{2j-1} \cdot {\bfm S}_{2j}. 
\end{eqnarray}
Here the spin ${\bfm S}$ is defined by  \be {\bfm S}_j = \frac 1 2\sum_{\sigma \sigma'} \hat c^\dagger_{j\sigma} {\bm \sigma}_{\sigma \sigma'} \hat c_{j\sigma'},\ee
and ${\bm \sigma}$ is a vector of Pauli matrices. 

Since the Peierls-Hubbard Hamiltonian at large enough $U$ can be thought of as a spin-1/2 chain, adding an explicit spin-spin interaction simply contributes to the dimerization of that spin chain. 
At the same time, it is clear that at very large negative $J\ll -t^2/U$, spins will have a tendency to form a spin-1 moment out of two spins on sites $2j-1$ and $2j$, converting the spin-1/2 antiferromagnetic chain into a spin-1 antiferromagnetic chain. 
The spin-1 antiferromagnetic chain is known to be in the Haldane phase, which is gapped and has zero-energy edge excitations. 

If we switch off $J$ while keeping $\delta t<0$, there is likely to be no phase transition, and the Haldane state will smoothly evolve into a state of a weakly dimerized spin-1/2 chain with topological invariant $N_1=2$. 
This argument supports the notion that the topological invariant in the Haldane state of a Heisenberg spin-1 chain is also $N_1=2$. 
Thus, generically, we expect that the boundary states of spin-1 Heisenberg chains in the Haldane phase are due to the nonzero value of the topological invariant.

We confirm this qualitative picture numerically in the next section.

\section{DMRG analysis of the topological invariant}
\label{sec:dmrg}

\begin{figure}[b]
\includegraphics[width=0.4451\textwidth]{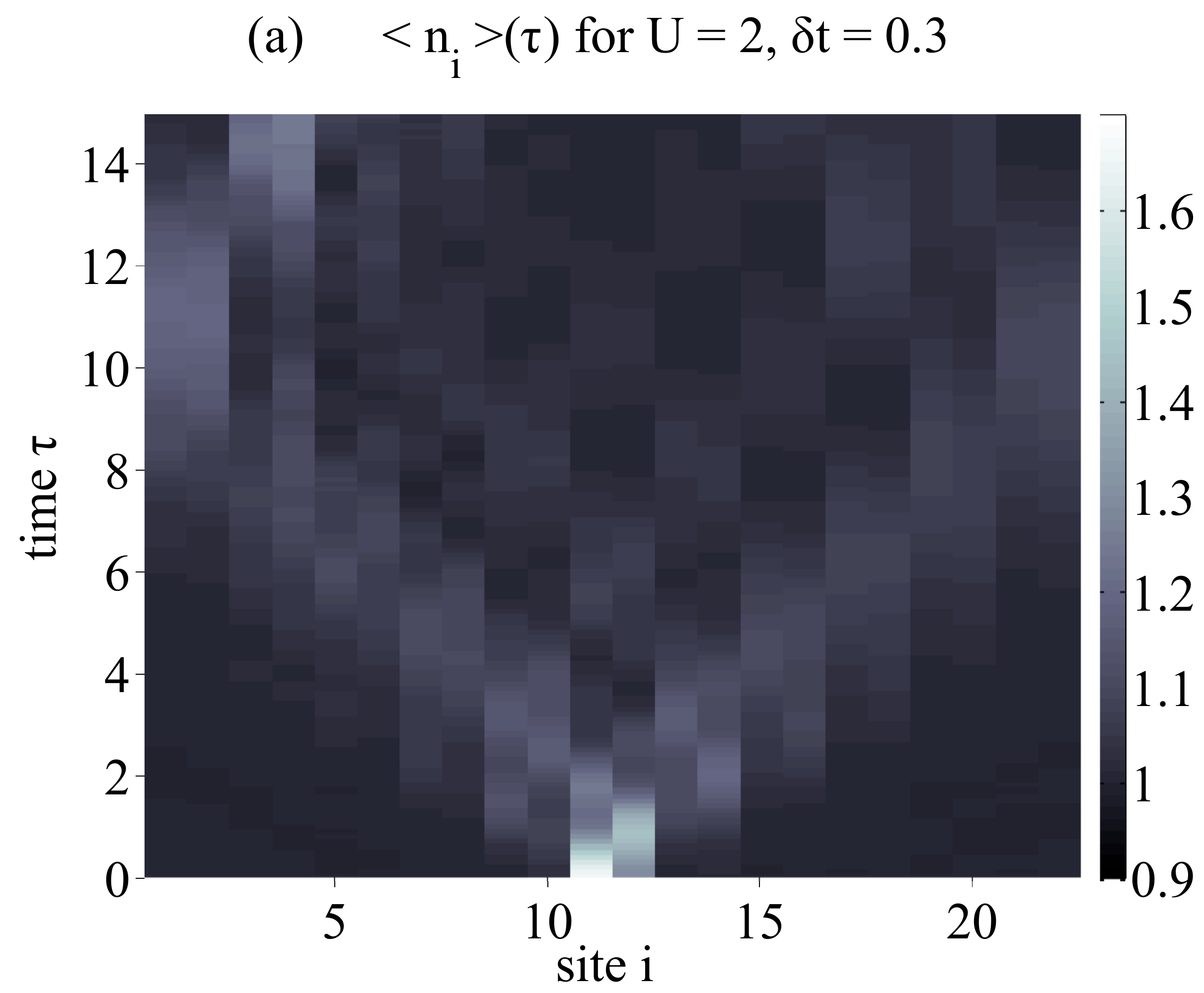}
\includegraphics[width=0.48\textwidth]{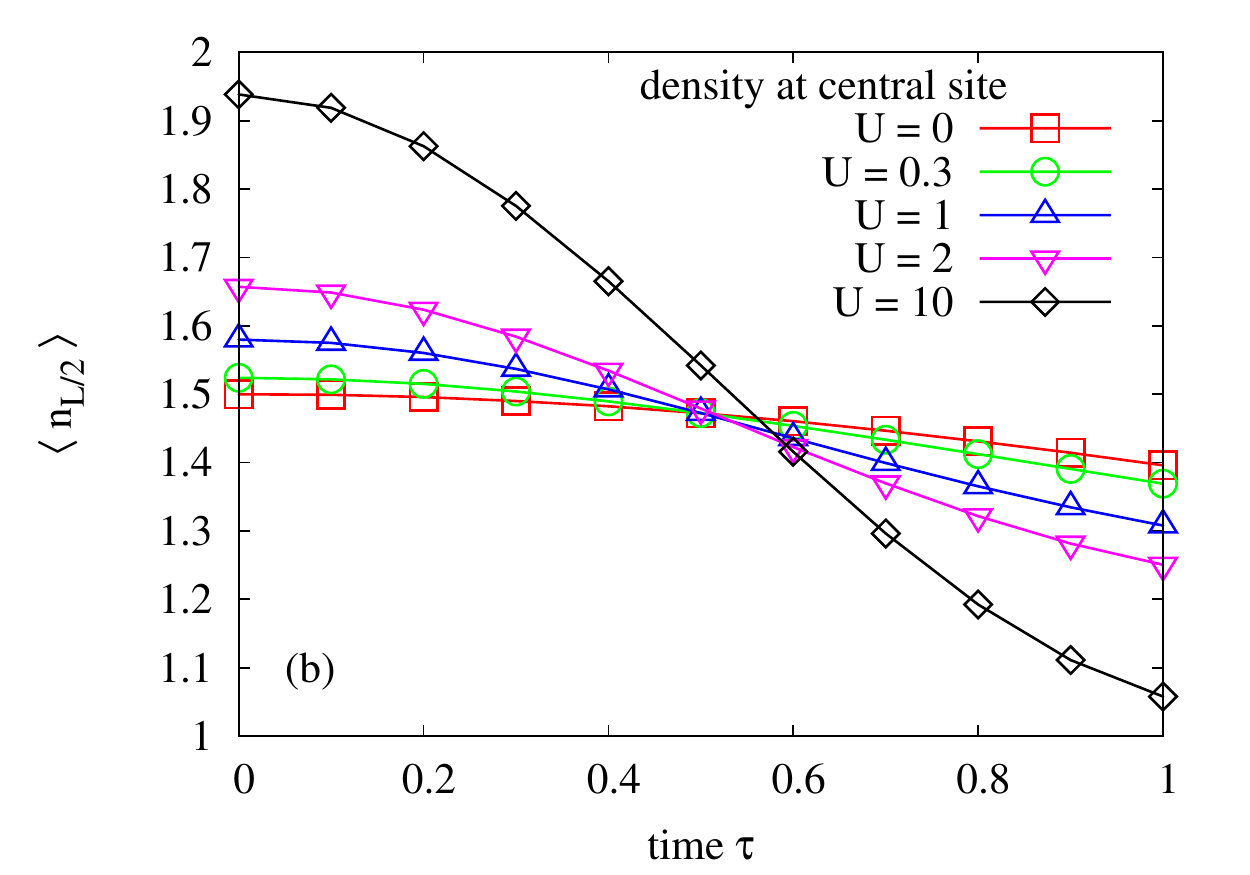}
\caption{(Color online) (a) Time evolution of the local density $\langle n_i \rangle(\tau)$ for a system with $L=22$ sites, $U/t=2$ and $\delta t = 0.3$. (b) Time evolution of the density on the central site for systems with $L=250$ sites, $\delta t = 0.3$, and for $U/t=0, \, 0.3, \, 1, \, 2, \, 10$.}
\label{fig:density}
\end{figure}

In this section, we discuss numerical results for the topological invariant in the Peierls-Hubbard model, Eq.~\eqref{eq:ham}, and for the extended model, Eq.~\eqref{eq:hamH}, in which the ferromagnetic couplings lead to a Haldane-like ground state.\cite{footnote}
We apply a Krylov variant\cite{Manmana:2005p63} of the adaptive t-DMRG\cite{Daley:2004p2943,White:2004p2941} to obtain the (real time) Green's function 
\begin{equation}
G(\tau,l) = \langle \hat c_l^{\phantom{\dagger}}(\tau) \, \hat c_{L/2}^{\dagger}(0)\rangle \, \theta(\tau) 
\label{eq:nonequaltimeG}
\end{equation}
(here $\theta$ is equal to $1$ if its argument is positive and $0$ otherwise) for systems with up to $L=250$ lattice sites.  To keep the notation simple, we suppress the spin indices on the fermion creation and annihilation operators, but the Green's function in Eq.~\rf{eq:nonequaltimeG} is calculated for, say, spin-up fermions. 
Note that here and in the following we use the symbol $\tau$ for the time in order to avoid confusion with the hopping strength $t$.  
To fix the units of energy and time, we set $t=1$ throughout the section (although for clarity we keep $t$ explicitly in some of the expressions below). 
We compute the topological invariant by obtaining the Fourier transform of this quantity, which we then use to analyze the expression in Eq.~\eqref{eq:top}.    
However, since the relevant information is encapsulated in the phase of the Green's function, we avoid  computing this quantity directly, which would require the computation of the numerical derivative $\partial_k \, G(k,\omega)$, and instead analyze the winding of the chiral phase, which we define as
\begin{equation}
V(k, \omega) = \arg \left[\int_0^\infty \! d\tau \, e^{i \omega \tau} \sum_{l\,\mathrm{even}} e^{i k (\frac l 2 - \frac {L+2} 4)} G(\tau,l) \right]. 
\label{eq:chiralwinding}
\end{equation}
Here, $l/2$ effectively  labels unit cells, $k$ ranges from $-\pi$ to $\pi$, and the formula is written assuming $L/2$ is odd.
The winding $V(k,\omega)$ computed at $\omega=0$  is nothing but the argument of $z(k)$ introduced in Eq.~\eqref{eq:winding}. [More precisely, $v(k)$ introduced there can be reduced to a number in our case, and $V(k,0)$ is its argument.]  

Calculating $V$ at $\omega=0$ directly is not easy due to the slow convergence of the $\tau$ integral, Eq.~(\ref{eq:chiralwinding}). 
Instead, we use the analyticity and continuity of $G$ to calculate its Fourier transform at a positive imaginary value of $\omega$, which ensures convergence.  
This does not introduce an error because the quantity we seek to calculate is an integer, and therefore is not sensitive to perturbations. 

Note that two complications limit the possible values of ${\rm Im}\,\omega$ that can be used for this procedure.  First, the magnitude of ${\rm Im}\,\omega$ must be such that, for the finite systems under consideration, times after the perturbation has reached the edges are suppressed. Hence ${\rm Im}\,\omega \gtrsim v/L$, with $v$ the speed of the fastest excitations in the system. 
Second, ${\rm Im}\,\omega$ should not be too large compared to typical energies of the system, since otherwise the structure of the Green's function will be determined by (the time cutoff provided by) ${\rm Im}\,\omega$ rather than by the eigenstates of the system. We conclude that ${\rm Im}\,\omega$ should be chosen to be of the order of the bandwidth, and, for the sake of simplicity, we choose $\omega = i$ for all cases treated in this paper, without a noticeable effect on the value of the topological invariant. 


As explained below, finite-size effects are of minor importance. 
Thus, in order to reduce the computational effort, most of the results presented are obtained for systems with only $L=22$ lattice sites.  
For these computations, we typically choose a time step $d\tau = 0.05$ and keep up to $m=500$ density matrix eigenstates in the course of the time evolution, leading to a discarded weight of the order of $10^{-5}$ at the end of the time evolution ($\tau=30$). 
We adapt the number of basis states by choosing a dynamical block state selection scheme (DBSS),\cite{dbss_legeza1} and fix the threshold for the quantum information loss (measured by the Kholevo bound, see Ref.~\onlinecite{dbss_legeza2} for details) to $\chi = 10^{-7}$.  
Note, however, that, as we discussed above, the main properties of the chiral phase are determined by the behavior at short times ($\tau \lesssim 5$), so that the accuracy for these results is higher. 

Given a chiral phase, we simply plot $V(k,i)$ as a function of $k$ and check if it winds around the unit circle as $k$ goes over the Brillouin zone, from $-\pi$ to $\pi$. If it does not wind, then $N_1=0$. If it winds once, $N_1=2$. (The invariant is twice the winding because of spin.)   

In addition to computing the chiral phase, Eq.~\eqref{eq:chiralwinding}, we also analyze results for the real time evolution of the Green's function, Eq.~\eqref{eq:nonequaltimeG}, for the local particle density $\langle n_i\rangle(\tau)$, and for the local spin density $\langle S^z_i \rangle(\tau)$, with a particular focus on the role of edge states in the course of the time evolution. 
Due to the mapping \eqref{eq:mappingnegativeU}, we treat only $U \geq 0$ and expect the same behavior for negative $U$, with charge and spin interchanged.  
We start our discussion of the numerical results by considering the local observables $\langle n_i \rangle(\tau)$ and $\langle S^z_i\rangle(\tau)$. 
In Fig.~\ref{fig:density}, we show a typical example for a system of $L=22$ sites at $U = 2, \, \delta t = 0.3$. 
As expected, at $\tau = 0$ the density away from the center is very close to $\langle n_i \rangle = 1$ and does not show any signatures of end states for either positive or negative values of $\delta t$. 
However, one realizes that due to the dimerization, the density distribution on the two central sites at the center of the system is unequal.
In addition, as shown in Fig.~\ref{fig:density}(b), adding a particle leads to a larger density on that site with increasing $U$. 
For $U=10$, near double occupancy is achieved. 
Interestingly, in the time evolution of the density at the central site [Fig.~\ref{fig:density}(b)] the curves seem to intersect at $\tau \approx 0.5$ for values of $U$ small enough. 

\begin{figure}[b]
\includegraphics[width=0.4451\textwidth]{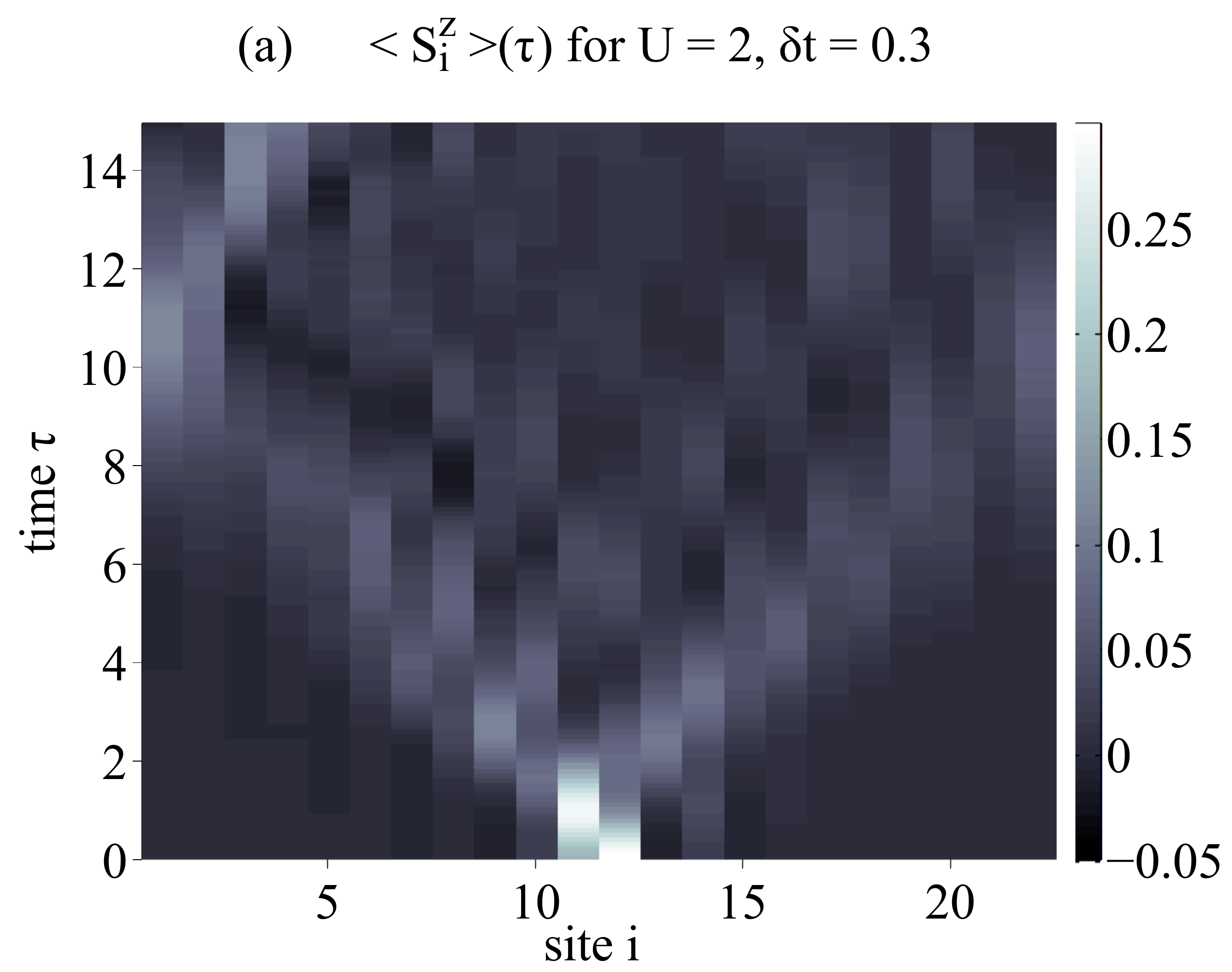}
\includegraphics[width=0.4451\textwidth]{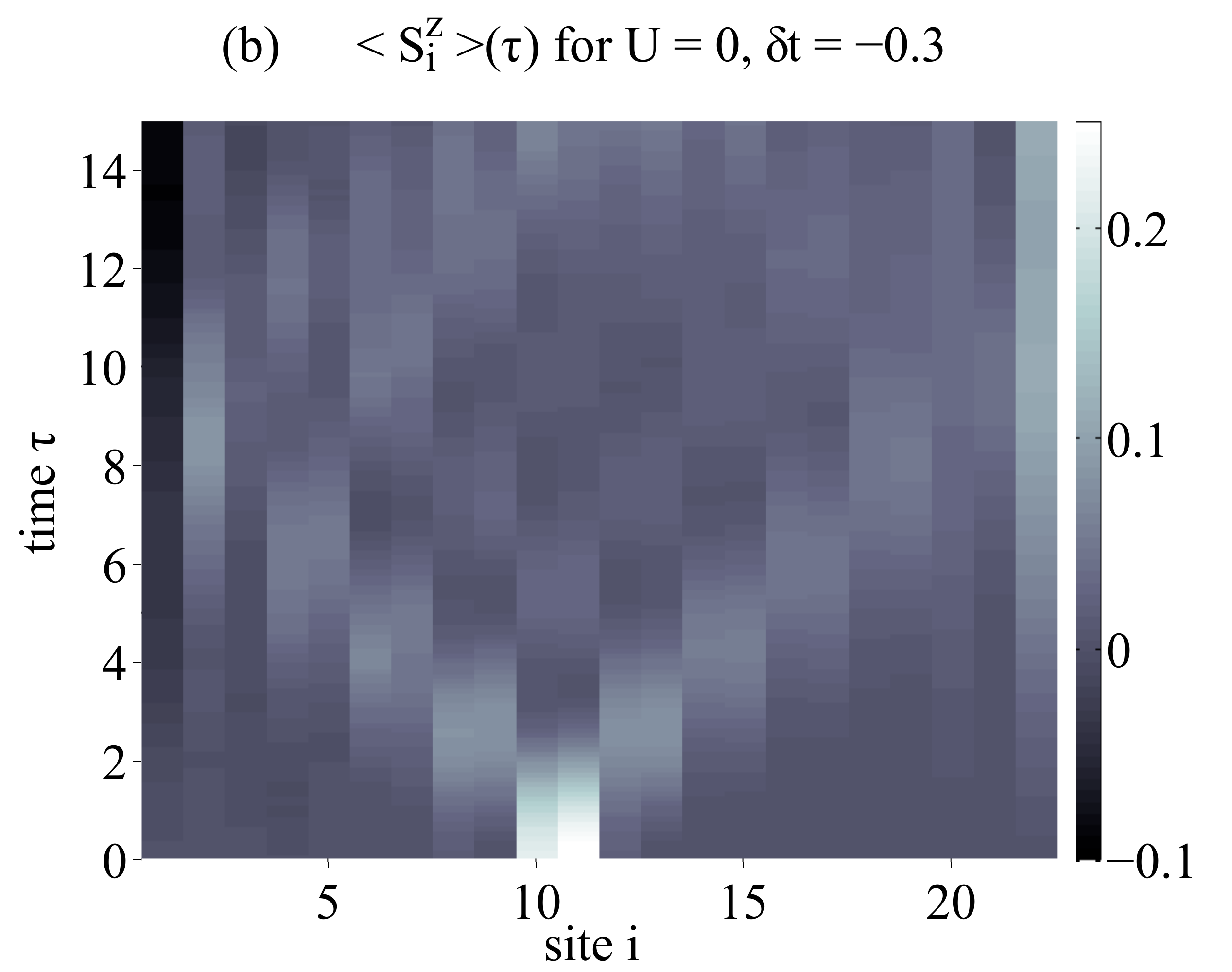}
\includegraphics[width=0.4451\textwidth]{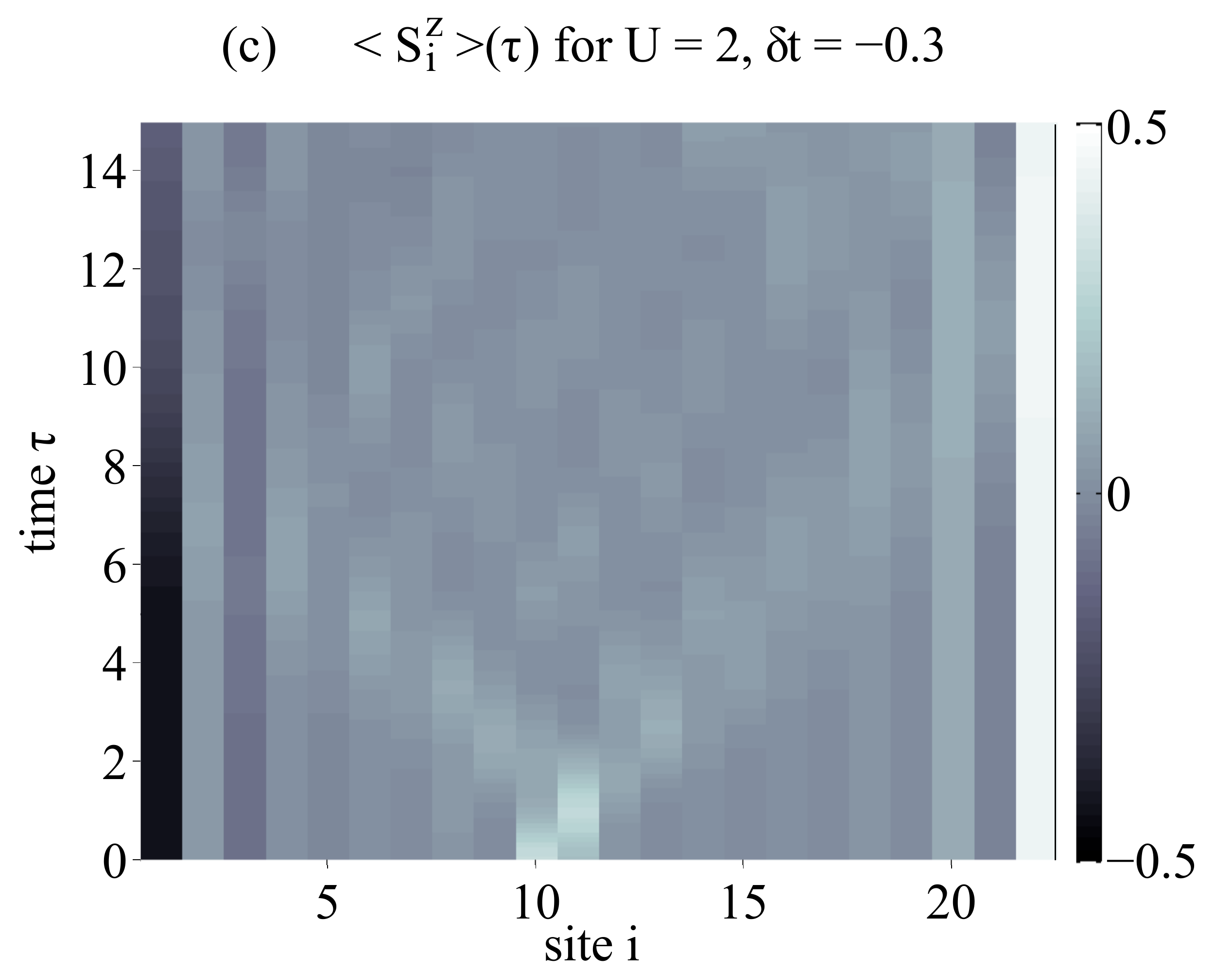}
\caption{Time evolution of the local spin density $\langle S^z_i \rangle (\tau)$ for (a) $U = 2$ and $\delta t = 0.3$, (b) $U = 0, \, \delta t = -0.3$, and (c) $U = 2, \, \delta t = -0.3$.}
\label{fig:sz}
\end{figure}

Edge states appear in the time evolution of the local spin density $\langle S^z_i\rangle(\tau)$. 
As can be seen in Fig.~\ref{fig:sz}, for positive values of $\delta t$, we find that $\langle S^z_i\rangle(\tau=0) \approx 0$ away from the central sites.  
However, for negative values of $\delta t$,  boundary states appear. 
For small values of $U$, Fig.~\ref{fig:sz}(b), no signature of boundary states can be seen at the beginning of the time evolution. 
However, in the course of the time evolution, local spins (pointing in opposite directions) seem to emerge at the two boundaries.
For larger values of $U$ [Fig.~\ref{fig:sz}(c)], these end spins are already visible in the initial state and appear to be close to fully polarized, again with opposite orientations at the two boundaries. 
This is consistent with the picture of boundary states for $\delta t<0$, a picture we subsequently confirm with the computation of the topological invariant. 

\begin{figure}[b]
\includegraphics[width=0.48\textwidth]{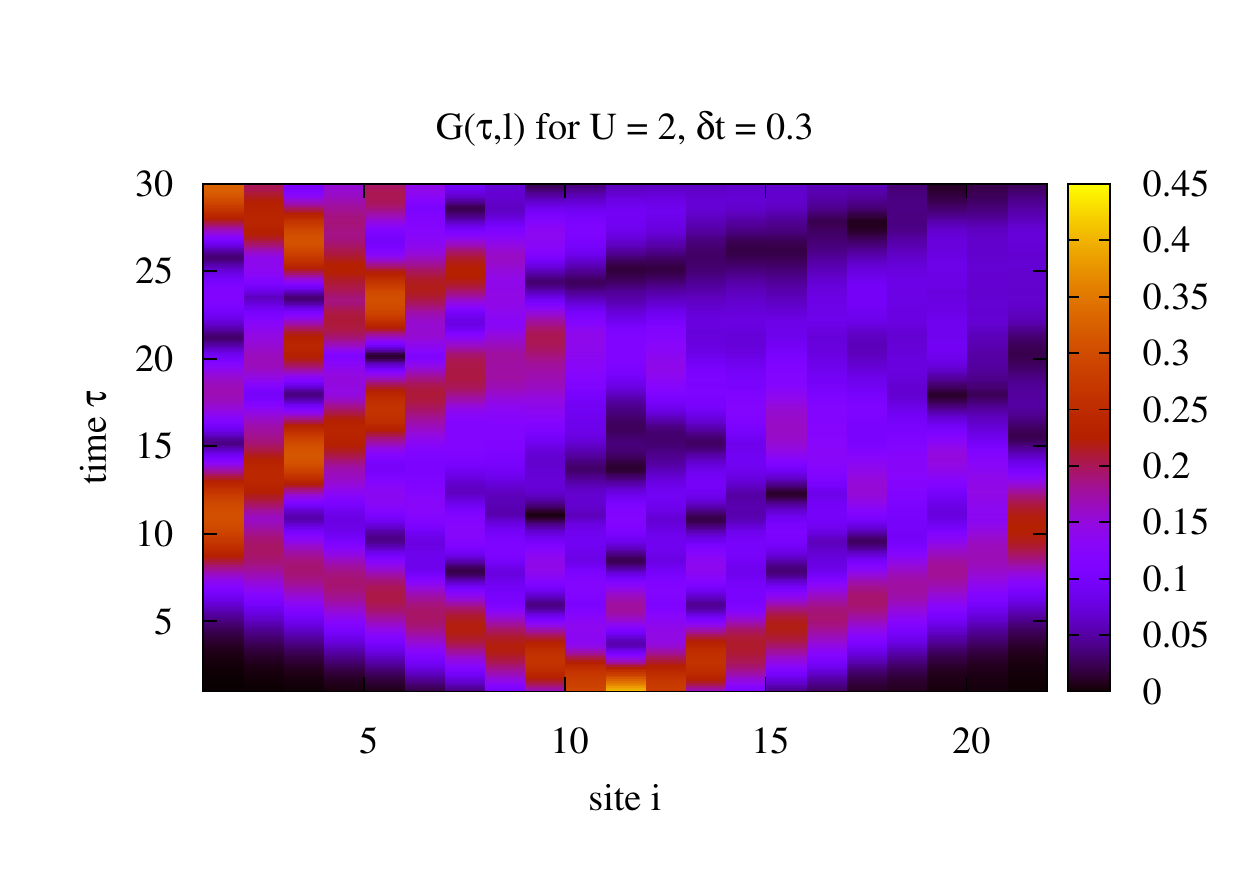}
\caption{(Color online) Time evolution of the Green's function for a system with $L=22$ sites with $U=2, \, \delta t = 0.3$. In the case shown, up to $m=750$ density matrix eigenstates were kept during the time evolution.}
\label{fig:greenfct}
\end{figure}

Now we turn to the time evolution of the Green's function. 
In Fig.~\ref{fig:greenfct}, we show typical results for the case of a system with $L=22$ sites, $U=2$, and $\delta t = 0.3$ for times up to $\tau = 30$, the maximum time reached. 
Again, the dimerization leads to an asymmetry in the initial state. 
The perturbation in the center spreads through the system with a typical velocity that depends on the values of $U$ and $\delta t$. 
No signature of boundary states can be seen for all values of $U$ and $\delta t$ treated. 
As can be seen, at a time $\tau \approx 10$, the perturbation reaches the boundary and gets reflected. 
This can lead to significant finite-size effects in the Fourier transform at $\omega = 0$, giving another reason why one should set  $\omega=i$, as discussed earlier. 
This effectively suppresses contributions to $V$ from late times, which are influenced by reflection from the boundaries, and so better captures the behavior of the infinite system.
We find that for $L=22$, all values of $U<10$, and $ - t \leq \delta t \leq t$, the perturbation reaches the boundary at a time $\tau \approx 5$ or later. 
By choosing $\omega = i$, which is a value of the same order of magnitude as the bandwidth, results for $G(\tau,l)$ at times $ \tau > 5$ have a very small weight ($< 1\%$) in the Fourier transform Eq.~\eqref{eq:chiralwinding}, so that finite-size effects should become minimal, as further discussed below.

\begin{figure}[t]
\includegraphics[width=0.48\textwidth]{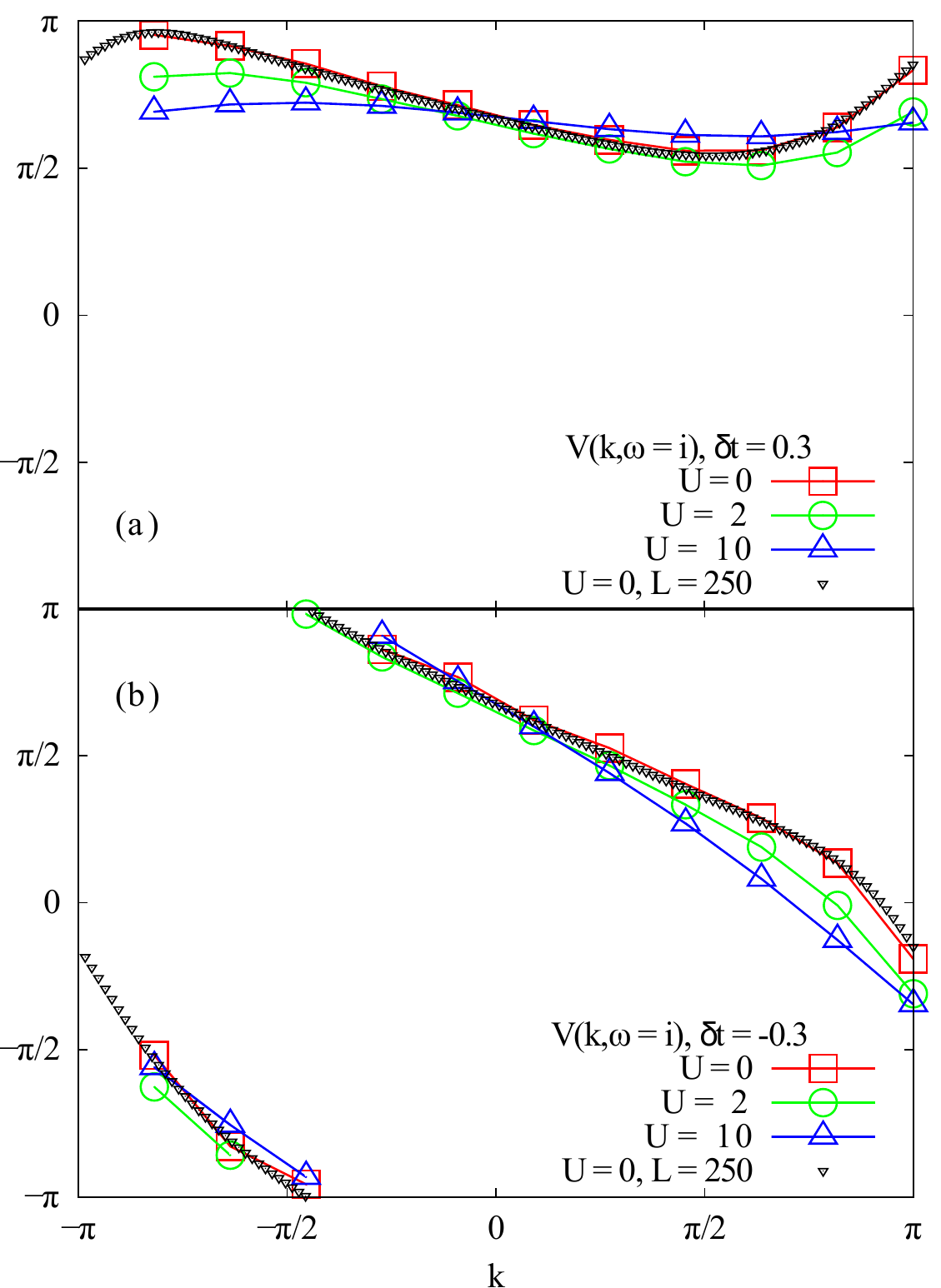}
\caption{(Color online) Chiral winding $V(k,i)$ [Eq.~\eqref{eq:chiralwinding}] for systems with $L=22$ sites (large symbols, solid lines) and $L=250$ sites (small symbols). (a) Results for $\delta t = 0.3$ and $U=0, \, 2, \, 10$. (b) Results for $\delta t = -0.3$ and the same values of $U$. }
\label{fig:winding}
\end{figure}

Based on this, we present in Fig.~\ref{fig:winding} results for $V(k,\omega=i)$ for $\delta t = \pm 0.3$ and $U=0, \, 2, \, 10$.  
For positive values of $\delta t$, we do not expect boundary states to be present, and there should be no winding of $V(k,i)$, while for negative values of $\delta t$, one winding of $V(k,i)$ should be obtained. 
This is indeed the case: 
as can be seen in Fig.~\ref{fig:winding}, for positive as well as negative values of $\delta t$ the function $V(k,i)$ appears to be periodic.
As seen in Fig.~\ref{fig:winding}(a), for $\delta t > 0$, the values covered are restricted to a region $\pi/2 < V(k,i) < \pi$, so that the winding number is zero.
For negative values of $\delta t$, however, the values cover the full range from $-\pi$ to $\pi$, so that the winding is equal to 1. 
Thus, we arrive at the main result of this paper: even in the presence of interactions, the topological invariant is equal to zero or $2$ (one winding per spin), and its value reflects the presence of boundary states.

\begin{figure}[t]
\includegraphics[width=0.48\textwidth]{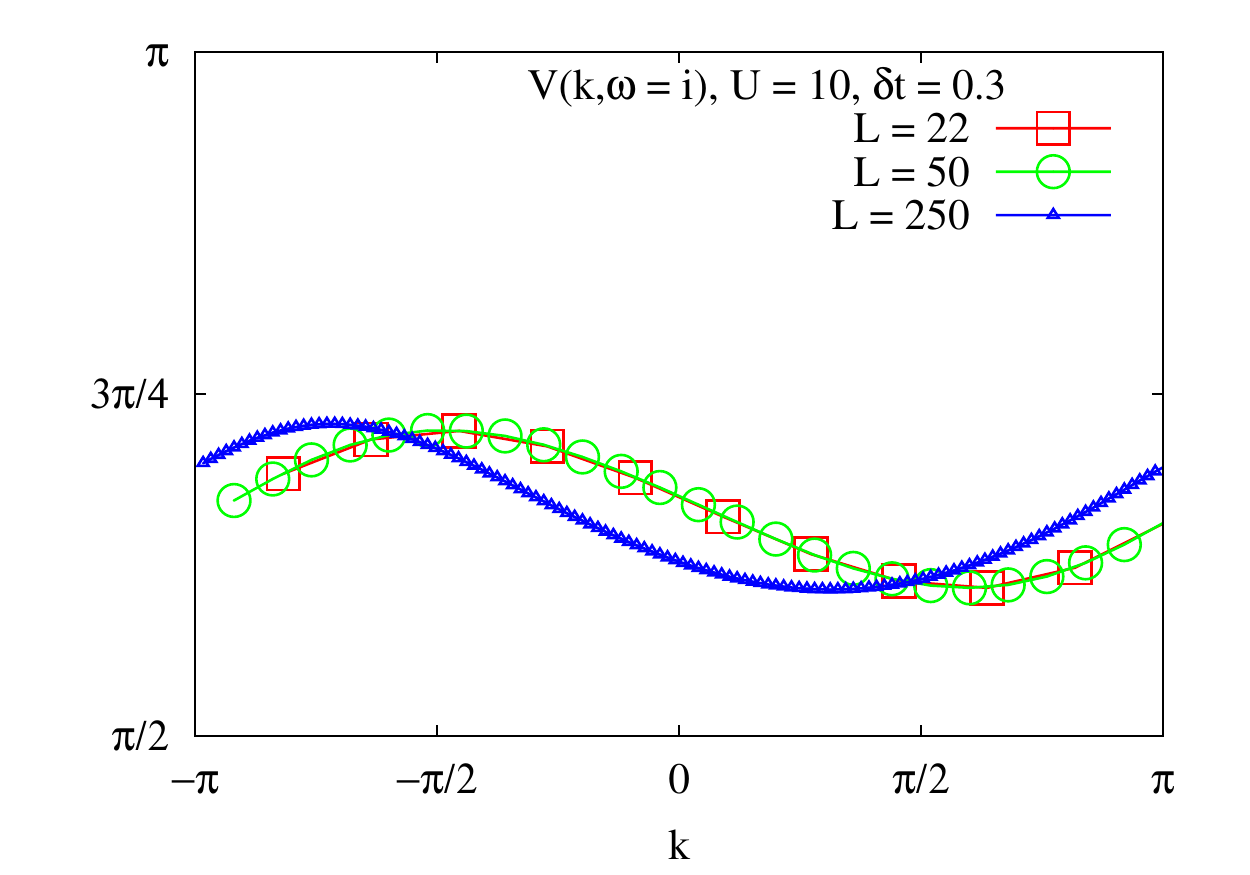}
\caption{(Color online) Chiral winding $V(k,i)$ at $U=10$ and $\delta t= 0.3$ for systems with $L = 22$, $L = 50$ and $L = 250$ lattice sites. While times $\tau = 12.5$ or larger were reached for the smaller systems, only times $\tau = 1.6$ were reached for $L=250$ sites.}
\label{fig:winding_differentL}
\end{figure}

\begin{figure}[b]
\includegraphics[width=0.48\textwidth]{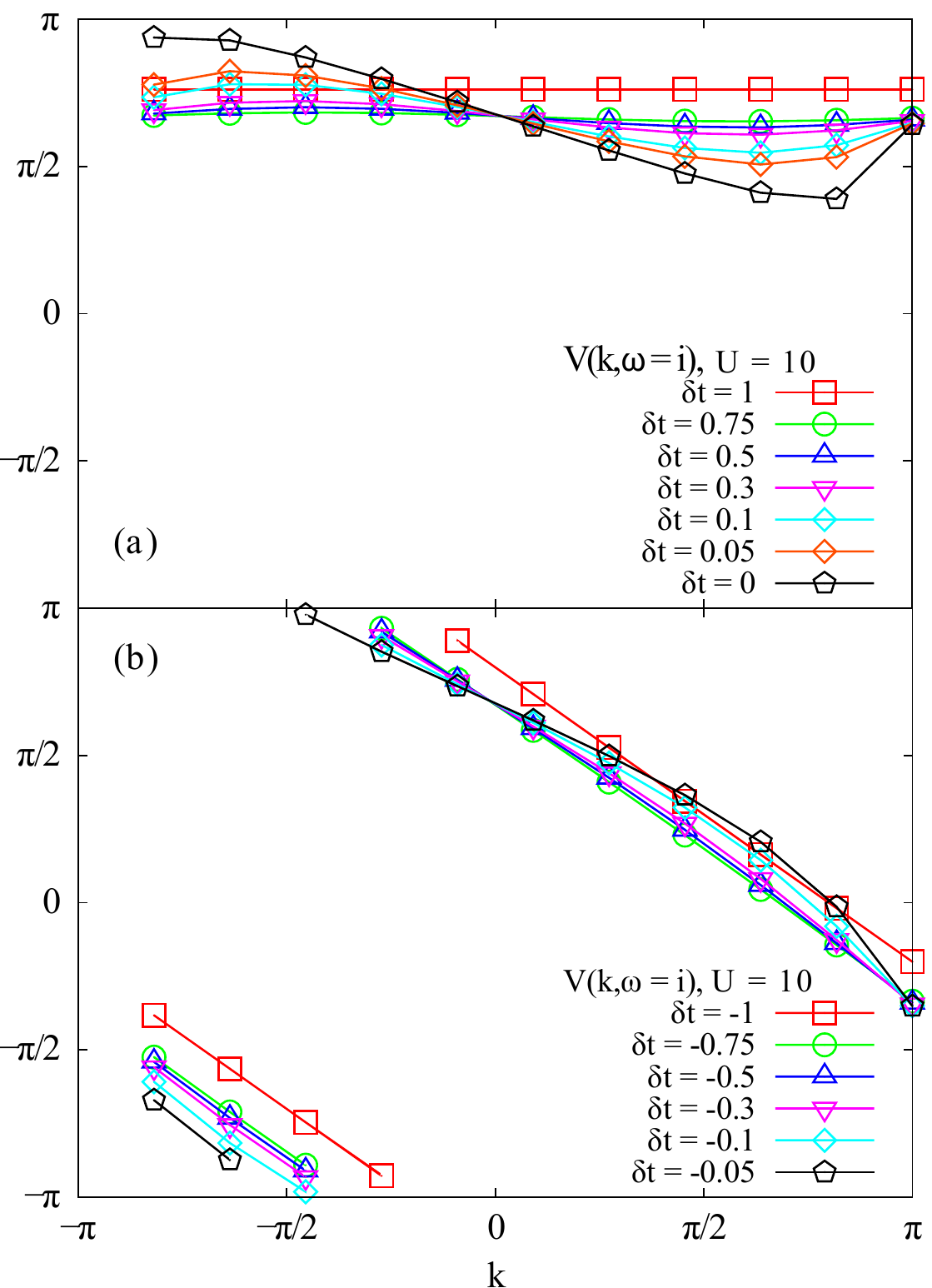}
\caption{(Color online) Chiral winding $V(k,i)$ for systems with $L=22$ sites, $U=10$ and different values of $\delta t$: (a) $\delta t \geq 0$; (b) $\delta t < 0$.}
\label{fig:winding_differentdelta}
\end{figure}

\begin{figure}[t]
\includegraphics[width=0.3\textwidth]{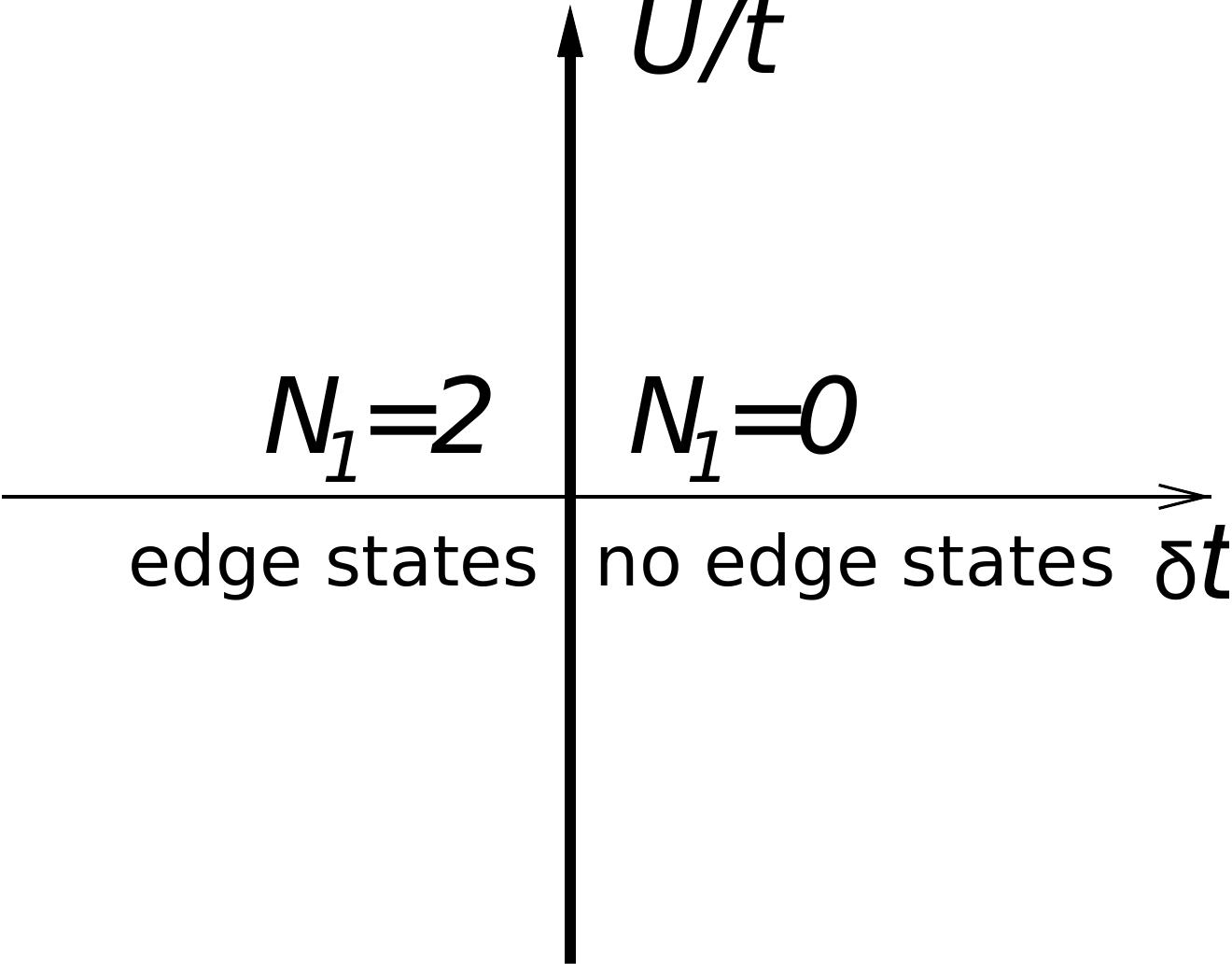}
\caption{Phase diagram of the Peierls-Hubbard model given by Eq.~(\ref{eq:ham}), with the value of the topological invariant $N_1$ indicated. The line $\delta t=0$ separates two gapped phases, topological and non-topological. The line $U=0$ describes the usual noninteracting 1D topological insulator. }
\label{fig:phasediag}
\end{figure}

Now we turn to finite-size and finite-time effects. 
In Fig.~\ref{fig:winding}, we compare our results for $V(k,i)$ for $U =0, \, \delta t = \pm 0.3$ to results with $L = 250$ sites. 
For $L=22$, times $\tau =15$ or larger were reached, so that, according to the above discussion, finite-time effects should not be a major issue. 
For $L=250$, the computations are more demanding and only times of the order of $\tau \approx 3$ were reached in the cases shown. 
Apparently, on the scale of the plot, finite-size and finite-time effects seem to be absent. 
We further analyze this in Fig.~\ref{fig:winding_differentL} in which we compare results for system sizes $L=22, \, 50$, and $L=250$ for $U=10$ and $\delta t = 0.3$. 
For $L=22$ and $L=50$, times of at least $\tau = 12.5$ were reached, while for $L=250$, only times of $\tau = 1.6$ were reached. 
By comparing the results for $L=22$ and $L=50$, we see that finite size effects seem to be practically absent also in this strongly interacting case. 
However, the results for $L=250$ seem to be shifted. 
We associate this with the small times that were reached. 
Importantly, this finite-time effect does not affect the overall behavior, and the winding number is correctly obtained. 
We therefore conclude that computing the chiral phase is, at least in the present case, a very stable numerical procedure  and can be performed with a rather moderate numerical effort for rather small systems and short times.  
Finally, we discuss results for fixed values of $U$ and varying $\delta t$, shown in Fig.~\ref{fig:winding_differentdelta} for $U=10$ and $L=22$. 
As can be seen, for all values of $\delta t \leq -0.05$ shown, one winding is present, while for all values of $\delta t \geq 0$ there is no winding. 
This is in agreement with the well-known phase diagram of the Peierls-Hubbard model,\cite{PhysRevB.66.045114} in which the spin gap closes on the line $(U, \delta t = 0)$, signifying a phase transition. 
Via computing the topological invariant, Fig.~\ref{fig:winding_differentdelta} shows that this phase transition connects a topologically trivial phase for $\delta t > 0$ to one with boundary states at $\delta t < 0$. 
The phase diagram of the Peierls-Hubbard model with the two phases identified by the value of the topological invariant is shown in Fig.~\ref{fig:phasediag}.

\begin{figure}[b]
\includegraphics[width=0.48\textwidth]{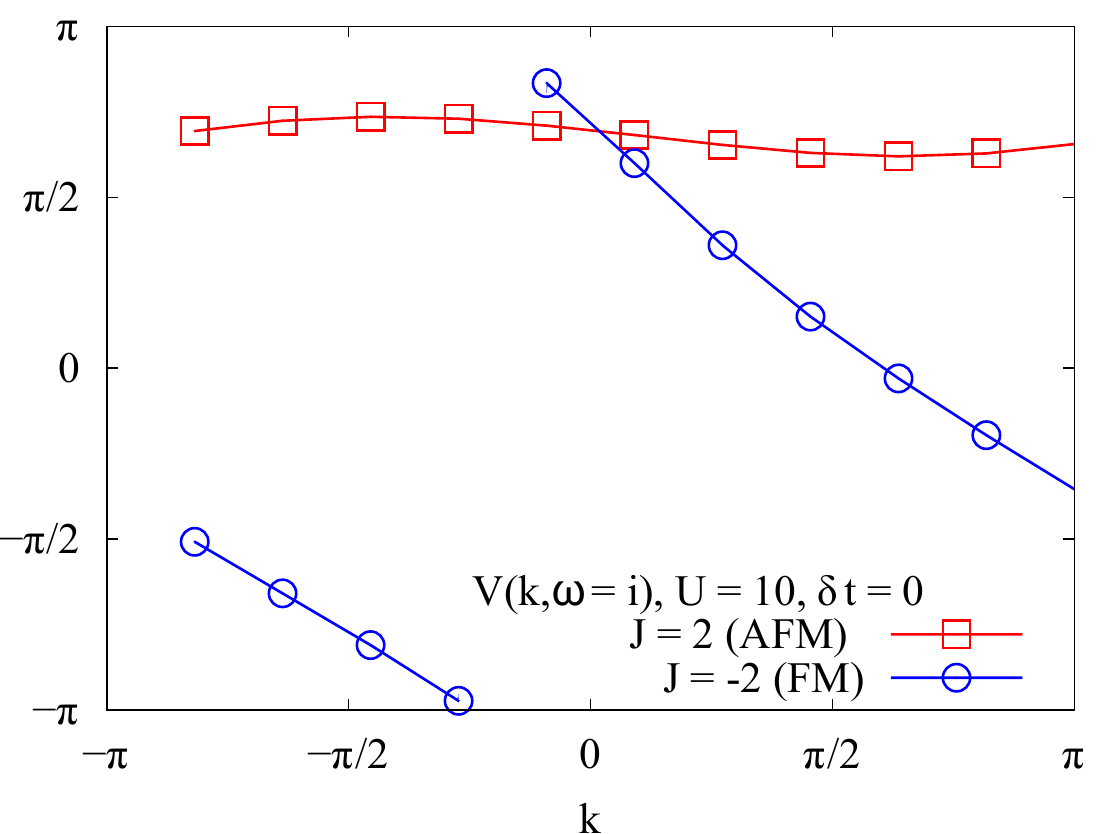}
\caption{(Color online) Chiral winding $V(k,i)$ for the extended model with spin-exchange terms, Eq.~\eqref{eq:hamH}, for systems with $L=22$ sites, $U=10$, and $J = 2$ (antiferromagnetic case, red squares) and $J=-2$ (ferromagnetic case, blue circles).}
\label{fig:resultsJterm}
\end{figure}

Based on these insights obtained in the Peierls-Hubbard model, we can now attempt to apply a similar procedure to other systems. 
Here, we discuss the chiral phase for the extended model \eqref{eq:hamH} with spin-exchange interactions. 
At $U>0$, the chain is a Mott insulator and, just as in the discussion of the model \eqref{eq:ham}, all its physics is in the interaction between the on-site spins. 
For sufficiently large ferromagnetic coupling $J/t\lesssim -t^2/U$, we expect the formation of effective spin-1 objects, leading to a ground state similar to that of a spin-1 Heisenberg chain with boundary states.
In contrast, antiferromagnetic coupling enforces the formation of singlets, so that the situation should be similar to the above discussion for the Peierls-Hubbard model at $\delta t > 0$. 
In Fig.~\ref{fig:resultsJterm}, we present results for $V(k,i)$ in the strongly interacting case at $U = 10$ with $L=22$ lattice sites for a ferromagnetic $J = -2$ and an antiferromagnetic $J=2$ with $\delta t =0$.  
We observe a winding of $1$ (corresponding, as before, to the topological invariant $N_1=2$) if $J=-2$  and $N_1=0$ if $J=2$. (Note that $t^2/U$ is $1/10$ and is much smaller than $J$.)  
This confirms that for negative $J$, where we expect this system to be a spin-1 Heisenberg chain, boundary states are present. 

We could now investigate other aspects of this system, such as the critical value of $J$ at which the invariant changes and the system goes through a quantum phase transition or the dependence of that transition on $\delta t$. 
However, since this would lead us too far 
afield, we leave the investigation of the full phase diagram of the model Eq.~\eqref{eq:hamH} to future studies.

\section{Analysis of the Fidkowski-Kitaev model}
\label{sec:fk}
As discussed in the introduction, a one-dimensional interacting system with the topological invariant that is a multiple of $4$ may have no boundary states whatsoever. 
This means that two Hamiltonians that are topologically distinct at the quadratic, noninteracting level can be adiabatically connected by adding an appropriate interaction.  
Here we demonstrate that such a system  must have Green's-function zeros at its boundary. (Correspondingly, zeros develop at  zero energy somewhere along the path in the parameter space that deforms one such system into another one with a distinct topological invariant.)

To carry out the demonstration, we present an analysis of the Fidkowski-Kitaev model,\cite{Fidkowski2010} which consists of two Peierls-Hubbard chains coupled by a Heisenberg interaction.  
The topological invariant of this model is either $0$ or $4$, even though it has only one phase. 
The Fidkowski-Kitaev Hamiltonian has a large symmetry that is more apparent when it is rewritten in terms of real (Majorana) fermion modes, which will be largely hidden in the treatment that follows. 

\begin{figure}[t]
\includegraphics[width=\columnwidth]{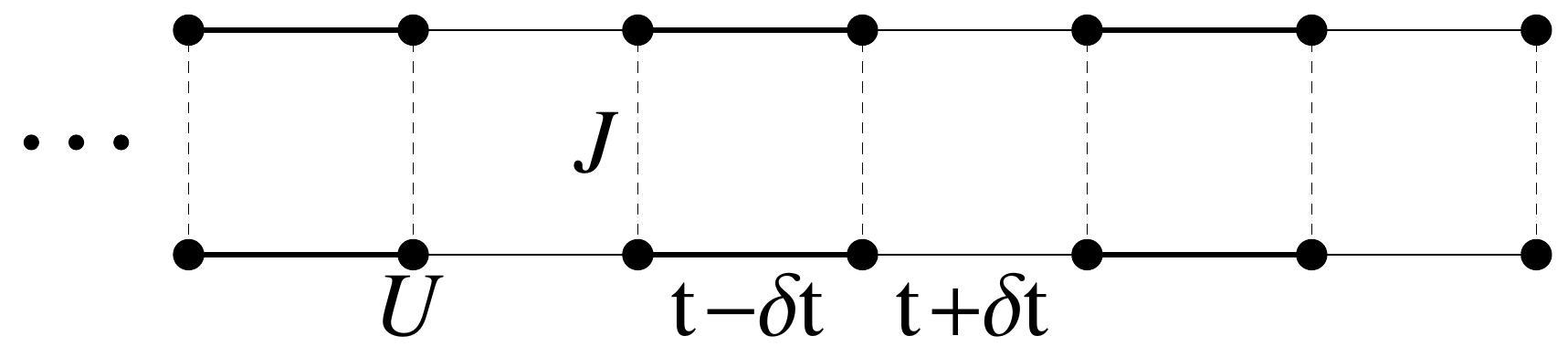}
\caption{\label{fig:ladder}
Graphical representation of the Fidkowski-Kitaev model of Ref.~\onlinecite{Fidkowski2010}: Two dimerized Hubbard chains are coupled via a ``rung'' Heisenberg exchange term.}
\end{figure}

The Heisenberg interaction between the two chains $a$ and $b$, described by fermion operators $\hat{a}_{i\sigma}$ and $\hat{b}_{i\sigma}$, is 
\begin{align}
H_{\mathrm{spin}} = J \bfm{S}_a \cdot \bfm{S}_b ,
\end{align}
where the spin operators are defined in terms of the fermion operators as before.  
We assume $J \ge 0$ (i.e., is antiferromagnetic) in the following, so that ``rung singlet'' is energetically favorable. 
The full Hamiltonian is depicted graphically in Figure~\ref{fig:ladder}.

The addition of the Heisenberg term allows a smooth interpolation between the two topologically distinct phases with $\mathrm{sign}\, \delta t = \pm \mathrm{sign}\, t$.  
Essentially, it provides an extra source of dimerization in the system that is lost when $\delta t = 0$, thereby keeping the spin waves gapped (in the case $U>0$). 
Let us outline the argument of Fidkowski and Kitaev to this effect and then reconcile this with the topological invariants of the single-particle Green's function.

As before, the simplest limiting case of the system is $\delta t = \pm t$.  
In this case, the topologically nontrivial phase for $J=0$ has zero-energy states located purely at the ends of the chain, with no tunneling.  
As we have seen, this is true even for nonzero $U$, when a single-particle gap opens but two degenerate ground states per chain, with a spin-1/2 at the end and no dynamics, remain.  
The Heisenberg term then causes a singlet to form from these spins, with a gap to the triplet excitations.

The bulk part of the fully dimerized chain is gapped, and adding the Heisenberg interchain interaction does not change  much qualitatively.  
After all, the dimers of the Hubbard chain can effectively be described by a Heisenberg interaction with $J' \sim t^2/U$.  Adding the extra Heisenberg term allows the dimerization to interpolate smoothly from the chains to the rungs.  
This means that, for $J/t>0$ and $U/t>0$, the tunneling can be turned off entirely without closing the gap, making the adiabatic continuation between the two noninteracting phases, which is confirmed by the lack of zero-energy modes at the ends, possible. 

The question is how can this be consistent with the fact that the noninteracting phases and the interacting phases with $J=0$ are distinguished by a topological  invariant?  
As before, the answer is that the bulk single-particle Green's function must develop zero-frequency zeros when $\delta t = 0$.  
Similarly, the Green's function for the end sites must have zero-frequency zeros for $\delta t < 0$.  
This situation is distinct from that of a single Hubbard chain, in which the Green's function breaks chiral symmetry due
to the degenerate ground state; here the ground state is unique.

To compute the Green's function, we must determine the ground state, the spectrum, and the matrix elements.  
We will only consider the Hamiltonian $H_{\mathrm{Hub}} + H_{\mathrm{spin}}$; it describes the end state of the chain with $\delta t = -t$ or the bulk for $t = \delta t = 0$. 
When $J>0$ the ground state, with energy $E_\mathrm{gs} = -3J/4-U$, is
\beq
\ket{\mathrm{gs}} = \frac{1}{\sqrt{2}}
\left( a^\dag_\up b^\dag_\down - a^\dag_\down b^\dag_\up \right) \ket{0},
\eeq
where $\ket{0}$ is the Fock vacuum of the $a$ and $b$ fermions.  The 
single-particle excitations above this ground state, given by acting on the ground state with a
single creation or annihilation operator, are all degenerate,
with energy $E_1 = -U/2$; there are eight such states.
The remaining states have energies $J/4-U$ (triplet) and $0$.  Tuning to the
point $J/4=U$ increases the symmetry of the model, as discussed at length
by Fidkowski and Kitaev.

The single-particle Green's function of the decoupled chain is given by
\begin{align}
G_{ij}(\omega) &= 
\me{\mathrm{gs}}{f_i (i\omega + E_\mathrm{gs} - H_{\mathrm{int}} )^{-1} f_j^\dag}{\mathrm{gs}} \notag\\
&\quad{}+ \me{\mathrm{gs}}{f_j^\dag (i\omega - E_\mathrm{gs} + H_{\mathrm{int}} )^{-1} f_i}{\mathrm{gs}}.
\end{align}
Here the fermion operator $f_i$ takes on the four values $a_\up$, $a_\down$, 
$b_\up$, and $b_\down$.  Given the properties listed above, the Green's 
function evaluates immediately to
\beq
G_{ij}(\omega) = \left[ \frac{1}{i\omega - 3J/4 - U/2} 
+ \frac{1}{i\omega + 3J/4 + U/2} \right] \delta_{ij}.
\eeq
This function satisfies chiral symmetry [\rfs{eq:chi}], but does not have a pole at zero frequency because the single-particle excitations are gapped.  
Instead, it has a zero, as $G_{ij}(0)=0$.  This confirms the reasoning outlined above.

\section{Discussion and conclusions}
\label{sec:conclusions}
In this paper, we have demonstrated the utility of calculating the topological invariant for interacting topological gapped systems, working with the example of spinful fermions hopping on a one-dimensional lattice. While the invariant is no longer directly related to conductivity or other responses of the system, it can still be used to deduce whether zero-energy boundary states are present, thanks to the bulk-boundary correspondence, Eq.~(\ref{eq:bb}). The invariant can be computed numerically; in this paper we accomplish this with the DMRG method. One advantage of computing topological quantities numerically is that, as integers, they are not strongly prone to numerical errors. 

One could, in principle, ask whether direct numerical evaluation of
the boundary states is no more difficult than evaluating the
topological invariant, or whether it may even be advantageous. 
We would like to point out that direct evaluation of this sort is more prone to numerical errors. 
The boundary states are susceptible to finite-size effects. 
While they could be lifted away from zero numerically, determining whether they are topologically protected might then not be easy. 
The topological invariant is robust and is only weakly susceptible to numerical errors. 
If it is found to be nonzero, the zero-energy boundary states are guaranteed to exist in the large-size limit. 

While direct numerical determination of the topological invariant such as Eq.~(\ref{eq:top}) can be  problematic due to the numerical errors associated with integrating derivatives of functions determined numerically, in the present one-dimensional case this was not necessary. Instead, we evaluated the winding associated with the topological invariants by inspection. For completeness, let us note that, if desired, we could have evaluated the invariant without having to inspect the graph of the chiral phase, Eq.~(\ref{eq:chiralwinding}), visually. 
Instead, we could have found all the solutions $k_i$ of the equation
\be \label{eq:v0} V(k_i,0) = V_0, 
\ee where $V_0$ is an arbitrarily chosen number between $-\pi$ and $\pi$. Given the set of $k_i$ which solves this, we can compute
\be \label{eq:numeval} N_1 = 2 \left( \sum_i {\rm sign} \, \left[ \left. \pbyp{V(k,0)}{k} \right|_{k=k_i}  \right] \right). 
\ee
This works for almost all $V_0$ and is $V_0$-independent. 
(It fails for those $V_0$ for which $V(k,0)$ has a vanishing derivative at $k$ being equal to one of the $k_i$.) 
If the derivative is too small to determine its sign dependably, a different $V_0$ can be chosen. 
As elsewhere throughout the paper, the prefactor $2$ has to do with the two spin components of spinful fermions.

Importantly, the formula Eq.~(\ref{eq:numeval}) has a natural counterpart in higher dimensions. \cite{KitaevDiscussion} For example, in two spatial dimensions one might want to evaluate the winding of a matrix $G(k,\omega)$  given by
\be  \label{eq:chern}\frac{1}{24\pi^2} \sum_{\alpha, \beta, \gamma} \epsilon_{\alpha \beta \gamma} \, \tr \int d\omega d^2 k \, G^{-1} \d_{\alpha} G G^{-1} \d_{\beta} G  G^{-1} \d_{\gamma} G.
\ee
Here $\alpha$, $\beta$, and $\gamma$ are summed over $\omega$, $k_x$, and $k_y$. 
This is equivalent to computing the Chern number if there are no interactions\cite{Niu1985} and can be reduced to a two-dimensional Berry-curvature integral even with interactions.\cite{zhong1,zhong2}
Evaluating the derivatives and the  integrals  in Eq.~(\ref{eq:chern}) numerically is problematic. 
Instead, in the important case where $G$ is a $2 \times 2$ matrix, one can parametrize it by writing it as a sum over a unit matrix and the three Pauli matrices with coefficients $v_0$, $v_1$, $v_2$, $v_3$. 
Since the overall normalization is irrelevant, this corresponds to three parameters $(\Theta_1, \Theta_2, \Theta_3)$ that are functions of $(\omega, k_x, k_y)$. 
Then, given $\vec{\Theta}$,  
we could evaluate the Jacobian at some special value of $\vec{\Theta}$, an analog of $V_0$ in Eq.~(\ref{eq:v0}).  
The sum of the signs of the Jacobians so computed is equal to Eq.~(\ref{eq:chern}). 
This method works if $G$ is a $2 \times 2$ matrix. Its generalization to the case where $G$ is a larger matrix is not known to us, but should, in principle, exist.  

Now that we have established that the method of topological invariants is useful for studying interacting fermionic systems in one-dimensional space, it would be interesting to apply it to other interacting topological insulators. One possible direction of further research would be to study two- and three-dimensional topological interacting systems. 
It might also be interesting to further apply this method to other one-dimensional problems, for example, spin chains and ladders accessible to the DMRG. 
It also would be worthwhile to clarify the relationship between this method and recently discussed symmetry-protected topological orders in one-dimensional space. \cite{Wen2011}

\acknowledgements
VG is grateful to A. Kitaev for discussions concerning numerical evaluations of topological invariants and to the Aspen Center for Physics where part of this work was done. VG and SRM were supported by the NSF grant no. PHY-0904017. We acknowledge C. Mund and A. Mai for work on the code and M. Hermele for helpful discussions.



\begin{thebibliography}{10}
\bibitem{footnote}
Note that in the DMRG we fix the number of particles to be $L$ (i.e. we are at half filling) and replace the particle-hole symmetric expression $U\, \Sigma_{j}  \left( \hat n_{j,\uparrow} - \frac 1 2\right) \left( \hat n_{j,\downarrow} - \frac 12 \right)$ by $U\, \Sigma_{j} \hat n_{j,\uparrow}\hat n_{j,\downarrow}$. In addition, we use a convention in which $t \to -t$, which can be done without affecting the discussion of our results.
\end{thebibliography}

\end{document}